\documentclass[10pt,preprintnumbers,superscriptaddress,tightenlines,nofootinbib, eqsecnum]{revtex4-2}

\usepackage{amsmath}
\usepackage{amsfonts}
\usepackage{amssymb}
\usepackage{bm}
\usepackage[colorlinks]{hyperref}
\usepackage{mathrsfs}
\usepackage{graphicx}   
\usepackage{subcaption} 
\usepackage{empheq}
\usepackage[normalem]{ulem}
\usepackage{tensor}
\usepackage{comment}
\usepackage[usenames]{color}
\hypersetup{linkcolor=blue}
\hypersetup{urlcolor=blue}
\usepackage[capitalize]{cleveref}
\allowdisplaybreaks

\DeclareSymbolFontAlphabet{\mathrsfs}{rsfs}
\DeclareMathAlphabet{\mathcal}{OMS}{cmsy}{m}{n}
\DeclareMathOperator\Log{Log}

\newcommand{\beq}{\begin{equation}}
\newcommand{\eeq}{\end{equation}}
\newcommand{\nm}{\nonumber}
\newcommand{\dd}{\mathrm{d}}

\begin{document}
 
\title{Gravitational radiation from inspiralling compact binaries to N$^3$LO \\ [0.2cm]   in the Effective Field Theory approach}

\author{Loris Amalberti}\email{loris.amalberti@desy.de}
\affiliation{Deutsches Elektronen-Synchrotron DESY, Notkestrasse 85, 22607 Hamburg, Germany}
\affiliation{Institut für Physik, Humboldt-Universität zu Berlin, 12489 Berlin, Germany}

\author{Zixin Yang}\email{zixin.yang@desy.de}
\affiliation{Deutsches Elektronen-Synchrotron DESY, Notkestrasse 85, 22607 Hamburg, Germany}

\author{Rafael A. Porto}\email{rafael.porto@desy.de}
\affiliation{Deutsches Elektronen-Synchrotron DESY, Notkestrasse 85, 22607 Hamburg, Germany}

\date{\today}

\begin{abstract}
Within the context of the Effective Field Theory (EFT) framework to gravitational dynamics, we compute the Hamiltonian, source quadrupole moment, and gravitational-wave energy flux for (non-spinning) inspiralling compact binaries at next-to-next-to-next-to leading order (N$^3$LO) in the Post-Newtonian (PN) expansion. We use the recently developed $d$-dimensional multipole-expanded effective theory, and explicitly perform the matching to the (pseudo-) stress-energy tensor. The calculation involves Feynman integrals up to three- (conservative) and two-loop (radiative) orders, evaluated within dimensional regularization. Our (ambiguity-free) results confirm (for the first time) the value of the gravitational-wave flux for quasi-circular orbits at 3PN order, while paving the way forward to the inclusion of spin effects as well as higher order computations.\end{abstract}

\preprint{DESY-24-084}

\maketitle
\tableofcontents

\section{Introduction}\label{sec:intro}

The LIGO-Virgo-KAGRA collaboration has observed ${\cal O}(10^2)$ gravitational-wave (GW) signals from binary compact objects \cite{KAGRA:2021vkt}, including some that have been found, e.g.~\cite{Nitz:2021zwj,Mehta:2023zlk}, analyzing the publicly available data~\cite{KAGRA:2023pio}. Future detectors such as the Laser Interferometer Space Antenna (LISA)~\cite{amaroseoane2017laser}, Cosmic Explorer (CE)~\cite{Reitze:2019iox} and  the Einstein Telescope (ET)~\cite{Punturo:2010zz, Branchesi:2023mws}, are expected to significantly increase the detection rates. The sheer number of new sources that will be accessible to third-generation GW observatories thus highlights the necessity of establishing high-precision analytic waveform templates for binary searches, not only for detection but more importantly to extract accurate physical information regarding the wave's origin (see e.g.~\cite{LISAConsortiumWaveformWorkingGroup:2023arg} for the case of LISA sources).\vskip 4pt

In this paper we focus on the (weak-field and slow-velocity) Post-Newtonian (PN) expansion of the two-body problem in gravity, e.g. \cite{Blanchet:2013haa,Buonanno:2014aza,Schafer:2018jfw,Rothstein:2014sra,Porto:2016pyg,Goldberger:2022ebt}, and in particular the effective field theory (EFT) approach to gravitational dynamics~\cite{Goldberger:2004jt,Porto:2005ac,Porto:2006bt,Porto:2008jj,Porto:2008tb,Gilmore:2008gq,Goldberger:2009qd,Porto:2010tr,Porto:2010zg,Foffa:2011ub,Ross:2012fc}. In the realm of the PN approximation, the current state of the art for non-spinning bodies is the 4.5PN precision for the phase~\cite{Blanchet:2023bwj} (\emph{i.e.} the $(v/c)^9$ correction to the leading order), the 4PN order for both the GW flux and the dominant quadrupolar amplitude mode~\cite{Blanchet:2023sbv} and 3.5PN for the sub-leading ones~\cite{Faye:2012we,Henry:2021cek,Henry:2022ccf}. For the case of spinning bodies, on the other hand, the state of the art is at 4PN for the GW flux \cite{Cho:2021mqw,Cho:2022syn} and to 3.5PN order for the amplitude \cite{Porto:2012as,Henry:2022dzx,Henry:2023tka}, as well as 4.5PN in the radiation-reaction force \cite{Maia:2017gxn,Maia:2017yok}. Although, overall the EFT approach has achieved the most accurate description of the dynamics for spinning bodies \cite{Porto:2012as,Maia:2017gxn,Maia:2017yok,Cho:2021mqw,Cho:2022syn,Levi:2022rrq}, as well as in the {\it conservative} sector for non-spinning binaries, with partial results to next$^5$-to-leading order (N$^5$LO) \cite{Galley:2015kus,Porto:2017dgs,Foffa:2019yfl,Foffa:2019rdf,Blanchet:2019rjs,Almeida:2021xwn,Almeida:2023yia,Henry:2023sdy,Blumlein:2020pyo}, the computation of the GW energy flux has been performed only to N$^2$LO so far, namely at 2PN (and 4PN) order for non-spinning (spinning) compact objects, respectively~\cite{Leibovich:2019cxo,Cho:2021mqw,Cho:2022syn}. The purpose of this paper is to report on the derivation of the (non-spinning) GW flux at N$^3$LO within the EFT approach. As we shall see, our (ambiguity-free) results agree with what had been---up until only a year ago \cite{Blanchet:2023sbv}---one of the key ingredients for the state of the art in the modelling of binary compact objects in gravity \cite{Blanchet:2001aw,Blanchet:2004ek,Blanchet:2001ax},\footnote{To our knowledge, this is the first confirmation of the source GW flux obtained in \cite{Blanchet:2001aw,Blanchet:2004ek} at 3PN order, which had not been reproduced by an independent methodology until now.} thus paving the road forward to including spin effects to a similar N$^3$LO level of accuracy, as well as higher order computations.\vskip 4pt

There are several conceptual and computational issues that start to appear at N$^3$LO. Notably, the well-known ultraviolet (UV) divergences in both the equations of motion~\cite{Foffa:2011ub} and nonlinear radiative corrections~\cite{Goldberger:2009qd}. Dimensional regularization (dim. reg.), a method extensively employed in the EFT framework since the seminal work in~\cite{Goldberger:2004jt} (see also~\cite{Blanchet:2003gy,Blanchet:2005tk}), has become the weapon of choice to tackle divergent terms, appearing as poles $\propto (d-3)^{-1}$ (with $d$ the number of spatial dimensions), naturally yielding results which are devoid of the ambiguities that polluted the previous derivation \cite{Blanchet:2001aw,Blanchet:2001ax,Blanchet:2004ek}. While the ambiguity-free nature of the EFT approach, dealing with not only UV but also infrared (IR) divergences, has already been demonstrated for the conservative sector up to 4PN order~\cite{Galley:2015kus,Porto:2017dgs,Foffa:2019yfl}, the computation of the GW flux necessitates an equally careful examination of divergent terms as well as of the multipole expansion at the level of the action in $d$-dimensions, a subject only recently studied in~\cite{Amalberti:2023ohj}. After including all contributions form the source multipole moments, as well as hereditary effects \cite{Goldberger:2009qd}, the final result for the total radiated energy is devoid of singularities, as expected, and produces a radiated GW flux for quasi-circular orbits which is in perfect agreement with the original result in \cite{Blanchet:2001aw,Blanchet:2004ek}. The application of the steps described in this paper to incorporate spin effects up to 5PN order will be reported elsewhere.\vskip 4pt

This paper is organized as follows. A brief summary of the EFT approach is provided in \S\ref{sec:setup}. An independent derivation of the  conservative dynamics to N$^3$LO is offered in \S\ref{sec:cons}, while the contributions to the radiative multipole moments through a matching computation, the total radiated power, and radiated energy for quasi-circular orbits, are discussed in \S\ref{sec:rad}. We conclude in \S\ref{sec:concl}. Useful formulas are collected in App.~\ref{sec:app_2pn}, while other known contributions to the (lower order) multipoles are presented in App.~\ref{sec:app_3pn}, alongside various consistency checks.  All our computations are performed with the help of the software \texttt{Mathematica} and various associated packages, \texttt{FeynArts} \cite{Hahn:2000kx}, \texttt{xAct}~\cite{ACT} (including sub-packages \texttt{xTensor} and \texttt{xPert}), as well as the integration-by-parts (IBP) program \texttt{LiteRed}~\cite{Lee:2013mka}. A~computer-readable ancillary file detailing various results presented here is included with the submission of this paper.
\vskip 4pt
\textbf{Notation:} We use $\hbar=c=1$ units with the mostly negative metric convention. In the context of dim. reg., we work in $d+1$ spacetime dimensions, with one time and $d$ spatial dimensions. Greek letters denote Lorentz indices (running from 0 to $d$), and Latin letters the spatial ones (running from 1 to $d$).
Bold symbols denote spatial vectors, and we define the relative position, $\mathbf{r}\equiv\mathbf{x}_1-\mathbf{x}_2$, the relative velocity, $\mathbf{v}\equiv\mathbf{v}_1-\mathbf{v}_2$ and relative acceleration, $\mathbf{a}\equiv\mathbf{a}_1-\mathbf{a}_2$, respectively. We follow the standard definitions $M \equiv m_1+m_2$, $\nu \equiv m_1 m_2 /M^2$, $\Delta=(m_1-m_2)/M = \sqrt{1-4\nu}$, and $m_{Pl}\equiv 1/\sqrt{32 \pi G}$, for the total mass, symmetric mass ratio, mass difference, and Planck's mass, respectively. Finally, we use $\int \dd^{d}\mathbf{q}\,/(2 \pi)^{d}\equiv \int_{\mathbf{q}}$ for the integration measure, and follow the multi-index notation introduced in~\cite{Blanchet:1985sp}, \emph{i.e.} $x^L \equiv x^{i_1}x^{i_2}\dots x^{i_{\ell-1}}x^{i_{\ell}}$ and $I^L  \equiv I^{i_1i_2 \dots i_{\ell-1}i_\ell}$, for the coordinates and multipole moments, respectively.

\section{Effective field theory setup}\label{sec:setup}
In this section we give a brief overview of the EFT approach and point the reader to the reviews in \cite{Rothstein:2014sra,Porto:2016pyg,Goldberger:2022ebt}, as well as the previous N$^2$LO derivation in \cite{Leibovich:2019cxo}, for more details. The inspiral dynamics of compact binaries can be delineated across three distinct length scales: the characteristic size of the bodies, $r_s \sim GM$, the orbital separation between them, $r$, and the wavelength of the gravitational radiation, $\lambda_{\text{GW}}$. In the PN regime of small velocities, $v \ll 1$, these length scales form an intertwined hierarchical structure, 
\begin{equation}\label{eq:hierarchy}
    r_s \ll r \ll \lambda_{\text{GW}}\,,
\end{equation}
which enables us to disentangle distinct physical effects from different scales in terms of a unified expansion parameter,~$v$, thus facilitating the  computations of GW observables. After ``integrating-out" (solving-for) the short-distance scale of each compact body ($r_s$), the object is described in terms of a worldline point-particle action, $S_{\rm pp}$, which for our purposes can be written as $(a=1,2)$
\beq
S_{\rm pp} = \sum_a m_a \int d\tau_a + \cdots = \sum_a m_a \int \sqrt{ g_{\mu\nu}v_a^{\mu}v_a^{\nu} } \, d\sigma_a + \cdots\,,
\eeq
where $v_a^\mu \equiv \frac{dx_a^\mu}{d\sigma_a}$, with $\sigma_a$ an affine parameter. The ellipses include spin as well as higher order finite-size effects. In this paper we consider Einstein's gravity, described by the Einstein-Hilbert action,
\beq
S_{\rm EH}=-2 m^2_{Pl} \int\,\dd^4 x \sqrt{-g} \, R[g]\,,
\eeq
with $R[g]$ the Riemann curvature scalar. In the scenario in \eqref{eq:hierarchy}, one can split the gravitational field into separate, non-overlapping, {\it regions}, via \cite{Rothstein:2003mp,Goldberger:2004jt}
\begin{equation}\label{eq:regions}
    g_{\mu \nu} = \eta_{\mu \nu} + \frac{h_{\mu \nu}(x)}{m_{Pl}} = \eta_{\mu \nu} + \frac{\bar{h}_{\mu \nu}(x)}{m_{Pl}} +\frac{H_{\mu \nu}(x)}{m_{Pl}}\,,
\end{equation}
with $\eta_{\mu \nu}$ the Minkowski metric. The (off-shell) potential modes, responsible for the binding of the system, are denoted as $H_{\mu \nu}$, whereas the (on-shell) radiation modes, describing the GW emission, are $\bar{h}_{\mu \nu}$, respectively; obeying the following scaling rules \cite{Goldberger:2004jt}
\begin{equation}\label{eq:scaling_rules}
    \partial_0 H_{\mu \nu} \sim \left( \frac{v}{r} \right)\,, \quad \partial_i H_{\mu \nu} \sim \left( \frac{1}{r}\right)\,, \quad \partial_{\alpha} \bar{h}_{\mu \nu} \sim \left( \frac{v}{r}\right)\,.
\end{equation}
The dynamics of the system is obtained after integrating out both the potential and radiation modes in the effective theory, one scale at the time. For instance, up to 3PN order, the effective gravitational potential, $V[x_a]$, from which we can obtain the binding energy, may be obtained via 
\begin{equation}\label{eq:integrate_out}
    e^{-i\,\int dt\, V[x_a]}=\int \,D H_{\mu \nu} \,e^{i\,S_{\rm EH}\left[H\right]+i\,S^{(\bar h= 0)}_{\rm GF}\left[H\right]+i\,S_{\rm pp}\left[x_a,H\right]}\,,
\end{equation}
ignoring the radiation fields and using the standard harmonic gauge-fixing condition, $S^{(\bar h= 0)}_{\rm GF} \propto (\partial_{\alpha} H^{\alpha}_{\,\mu} -\frac{1}{2}\partial_{\mu}H^{\alpha}_{\, \alpha})^2$. For the sake of completeness, we derive the conservative effects at N$^3$LO \cite{Blanchet:2004ek}, first rederived within the EFT approach in \cite{Foffa:2011ub}, but we do so using different metric variables and gauge condition.\vskip 4pt 

For the derivation of radiative effects, on the other hand, we perform a matching computation. This is achieved by integrating out the potential modes in the {\it full theory}, including both $H$ and $\bar h$ fields, and reading off the (time-dependent) mass- ($I^L$) and current-type ($J^L$) couplings in the multipole-expanded effective theory \cite{Goldberger:2009qd,Ross:2012fc}, recently constructed in $d$ dimensions in \cite{Amalberti:2023ohj}. The long-distance effective action takes the form, schematically,\footnote{Formally speaking, the $J^L$ coupling in \eqref{effE} only exists in $d=3$ dimensions, and it must be extended to a generalized coupling to the curvature tensor for $d \neq 3$. Moreover, an extra (Weyl-type) multipole moment appears, which vanishes for $d=3$. (See \cite{Amalberti:2023ohj} for details.) Although, crucially, the $d$-dimensional expressions for the $(I^L,J^L)$ in terms of moments of the stress-energy tensor are needed, none of the subtleties associated with additional couplings play a role at N$^3$LO order, and first appear at 4PN \cite{Blanchet:2023bwj}.} \beq
S_{\rm eff} = \frac{1}{2} \int dt \big( I^L(t) \partial_{L-2} E_{i_{\ell-1}i_{\ell}} + J^L(t) \partial_{L-2} B_{i_{\ell-1}i_{\ell}}\big)\,,\label{effE}
\eeq
in terms of the electric, $E$, and magnetic, $B$, components of the Weyl tensor, which follows from a multipole expansion (around the binary's center-of-mass) of the linearized coupling,
\beq
\frac{1}{2 m_{Pl}} \int \dd^d x \, T^{\mu\nu} \bar h_{\mu\nu}\,,
\eeq
with $T^{\mu\nu}$ the (pseudo-)stress-energy tensor. See \cite{Goldberger:2009qd,Ross:2012fc, Amalberti:2023ohj} for details. Crucially, the matching computation is performed using a background-field gauge for the potential modes,
\begin{equation}\label{eq:gauge_fixing}
   S^{(\bar h)}_{\rm GF}=m^2_{Pl} \int \, \dd^4 x \sqrt{\bar{g}}\,\Gamma_{\mu}\Gamma^{\mu}\,, 
\end{equation}
where $\Gamma_{\mu} = D_{\alpha} H^{\alpha}_{\,\mu} -\frac{1}{2}D_{\mu}H^{\alpha}_{\, \alpha}$, using the covariant derivative associated to the (background) radiation field $\bar{g}_{\mu \nu}$. This ensures that the resulting $T^{\mu\nu}$ obeys the Ward identity, $\partial_\mu T^{\mu\nu} =0$, which is required for the construction of the (gauge-invariant) effective theory.

\section{Conservative dynamics at N$^3$LO}\label{sec:cons}

\begin{figure}[t!]
  \centering
  \begin{subfigure}[b]{0.13\textwidth}
    \includegraphics[width=\linewidth]{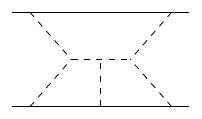}
  \end{subfigure}
  \begin{subfigure}[b]{0.13\textwidth}
    \includegraphics[width=\linewidth]{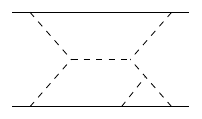}
  \end{subfigure}
  \begin{subfigure}[b]{0.13\textwidth}
    \includegraphics[width=\linewidth]{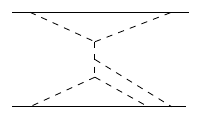}
  \end{subfigure}
  \begin{subfigure}[b]{0.13\textwidth}
    \includegraphics[width=\linewidth]{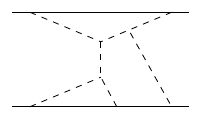}
  \end{subfigure}
  \begin{subfigure}[b]{0.13\textwidth}
    \includegraphics[width=\linewidth]{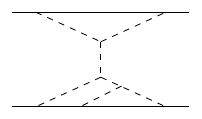}
  \end{subfigure}
    \begin{subfigure}[b]{0.13\textwidth}
    \includegraphics[width=\linewidth]{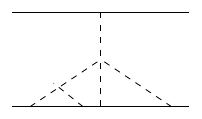}
  \end{subfigure}
      \begin{subfigure}[b]{0.13\textwidth}
    \includegraphics[width=\linewidth]{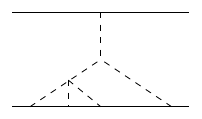}
  \end{subfigure}
      \begin{subfigure}[b]{0.13\textwidth}
    \includegraphics[width=\linewidth]{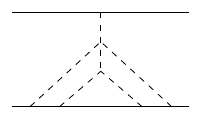}
  \end{subfigure}
      \begin{subfigure}[b]{0.13\textwidth}
    \includegraphics[width=\linewidth]{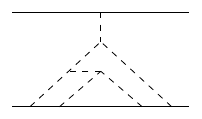}
  \end{subfigure}
      \begin{subfigure}[b]{0.13\textwidth}
    \includegraphics[width=\linewidth]{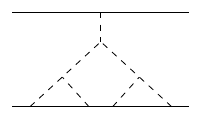}
  \end{subfigure}
    \begin{subfigure}[b]{0.13\textwidth}
    \includegraphics[width=\linewidth]{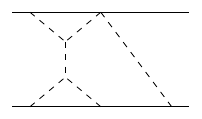}
  \end{subfigure}
  \begin{subfigure}[b]{0.13\textwidth}
    \includegraphics[width=\linewidth]{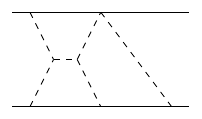}
  \end{subfigure}
  \begin{subfigure}[b]{0.13\textwidth}
    \includegraphics[width=\linewidth]{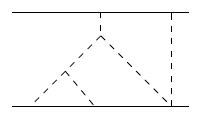}
  \end{subfigure}
  \begin{subfigure}[b]{0.13\textwidth}
    \includegraphics[width=\linewidth]{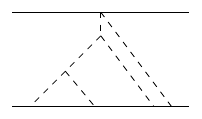}
  \end{subfigure}
  \begin{subfigure}[b]{0.13\textwidth}
    \includegraphics[width=\linewidth]{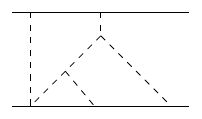}
  \end{subfigure}
    \begin{subfigure}[b]{0.13\textwidth}
    \includegraphics[width=\linewidth]{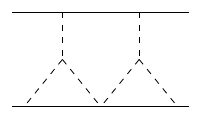}
  \end{subfigure}
  \begin{subfigure}[b]{0.13\textwidth}
    \includegraphics[width=\linewidth]{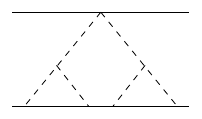}
  \end{subfigure}
  \begin{subfigure}[b]{0.13\textwidth}
    \includegraphics[width=\linewidth]{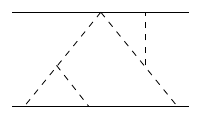}
  \end{subfigure}
  \begin{subfigure}[b]{0.13\textwidth}
    \includegraphics[width=\linewidth]{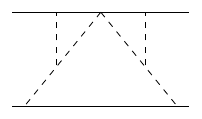}
  \end{subfigure}
  \begin{subfigure}[b]{0.13\textwidth}
    \includegraphics[width=\linewidth]{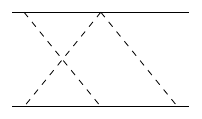}
  \end{subfigure}
    \begin{subfigure}[b]{0.13\textwidth}
    \includegraphics[width=\linewidth]{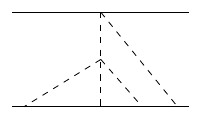}
  \end{subfigure}
  \begin{subfigure}[b]{0.13\textwidth}
    \includegraphics[width=\linewidth]{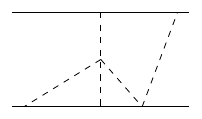}
  \end{subfigure}
  \begin{subfigure}[b]{0.13\textwidth}
    \includegraphics[width=\linewidth]{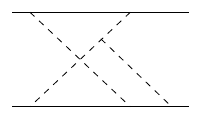}
  \end{subfigure}
  \begin{subfigure}[b]{0.13\textwidth}
    \includegraphics[width=\linewidth]{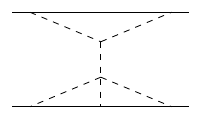}
  \end{subfigure}
  \begin{subfigure}[b]{0.13\textwidth}
    \includegraphics[width=\linewidth]{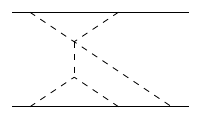}
  \end{subfigure}
    \begin{subfigure}[b]{0.13\textwidth}
    \includegraphics[width=\linewidth]{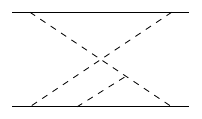}
  \end{subfigure}
  \begin{subfigure}[b]{0.13\textwidth}
    \includegraphics[width=\linewidth]{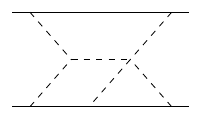}
  \end{subfigure}
  \begin{subfigure}[b]{0.13\textwidth}
    \includegraphics[width=\linewidth]{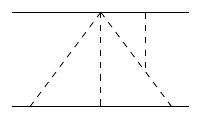}
  \end{subfigure}
  \begin{subfigure}[b]{0.13\textwidth}
    \includegraphics[width=\linewidth]{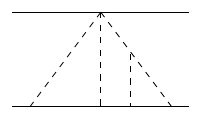}
  \end{subfigure}
  \begin{subfigure}[b]{0.13\textwidth}
    \includegraphics[width=\linewidth]{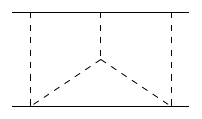}
  \end{subfigure}
    \begin{subfigure}[b]{0.13\textwidth}
    \includegraphics[width=\linewidth]{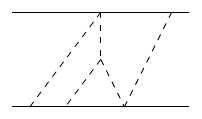}
  \end{subfigure}
  \begin{subfigure}[b]{0.13\textwidth}
    \includegraphics[width=\linewidth]{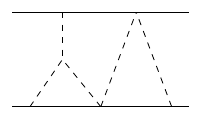}
  \end{subfigure}
  \begin{subfigure}[b]{0.13\textwidth}
    \includegraphics[width=\linewidth]{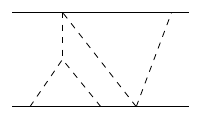}
  \end{subfigure}
  \begin{subfigure}[b]{0.13\textwidth}
    \includegraphics[width=\linewidth]{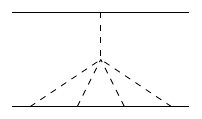}
  \end{subfigure}
  \begin{subfigure}[b]{0.13\textwidth}
    \includegraphics[width=\linewidth]{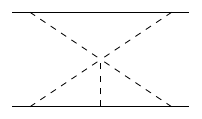}
  \end{subfigure}
    \begin{subfigure}[b]{0.13\textwidth}
    \includegraphics[width=\linewidth]{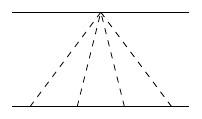}
  \end{subfigure}
    \begin{subfigure}[b]{0.13\textwidth}
    \includegraphics[width=\linewidth]{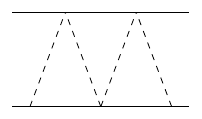}
  \end{subfigure}
  \begin{subfigure}[b]{0.13\textwidth}
    \includegraphics[width=\linewidth]{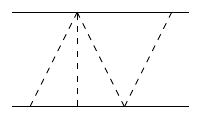}
  \end{subfigure}
  \caption{Topologies at  ${\cal O}(G^4)$ for the 3PN conservative dynamics (mirror images are omitted).}
  \label{fig:potential_3pn}
\end{figure}

We start by rederiving the conservative part of the dynamics of the binary system at 3PN order, which will be needed to derive the equations of motion entering in the GW flux. Moreover, since it is actually inconsequential in the radiation sector, we perform this computation without resorting to a further decomposition of the gravitational field into polarization modes \cite{Foffa:2011ub}, thus providing yet another independent check of the results. For the computation of the gravitational potential, we only require the (instantaneous) propagator for potential modes, with the structure,
\begin{equation}\label{eq:propagator}
 - i P_{\mu \nu; \alpha \beta}\, \delta(t)  \int_{\mathbf{q}} \, \frac{1}{\mathbf{q}^2} e^{i \mathbf{q}\cdot \mathbf{x}} \,,
\end{equation}
where $P_{\mu \nu; \alpha \beta}=\frac{1}{2}\left[\eta_{\mu \alpha}\eta_{\nu \beta}+\eta_{\mu \beta}\eta_{\nu \alpha}-\frac{2}{d-1}\eta_{\mu \nu}\eta_{\alpha \beta}\right]$.  We display in Fig.~\ref{fig:potential_3pn} a representative set of Feynman topologies, entering at ${\cal O}(G^4)$, with the dashed lines representing the propagator of potential modes and the solid lines representing the (non-propagating) worldlines, treated as external sources. See App.~\ref{sec:app_2pn} for additional terms, including those responsible for departures from instantaneity, see also App. \ref{app:sub_cons} for various other contributions.\vskip 4pt We employ the IBP program \texttt{LiteRed} to facilitate the reduction of the integrand families into a smaller collection of independent set of (``three-loop") Feynman master integrals, which we evaluate in dim. reg., yielding 
\begin{subequations}\label{eq:three_loop}
\begin{align}
M_{3,1}
& = 
\int_{\mathbf{q}_1,\mathbf{q}_2,\mathbf{q}_3}  \frac{1}{\mathbf{q}_1^2\left(\mathbf{q}_1-\mathbf{q}_2\right)^2\left(\mathbf{q}_2+\mathbf{q}_3\right)^2\left(\mathbf{q}_3-\mathbf{p}\right)^2}
= \frac{|\mathbf{p}|^{-8+3d}}{\left( 4 \pi \right)^{\frac{3 d}{2}}} \frac{\Gamma \left(4-\frac{3 d }{2}\right) \Gamma \left(\frac{d }{2}-1\right)^4}{\Gamma (2 d -4)} \,, \\
M_{3,2}
& = 
\int_{\mathbf{q}_1,\mathbf{q}_2,\mathbf{q}_3}  \frac{1}{\mathbf{q}_1^2\mathbf{q}_3^2\left(\mathbf{q}_1-\mathbf{q}_2\right)^2\left(\mathbf{q}_2+\mathbf{q}_3\right)^2\left(\mathbf{q}_2+\mathbf{p}\right)^2}
= \frac{|\mathbf{p}|^{-10+3d}}{\left( 4 \pi \right)^{\frac{3 d}{2}}} \frac{\Gamma (3-d ) \Gamma \left(2-\frac{d }{2}\right) \Gamma \left(\frac{d
   }{2}-1\right)^5 }{\Gamma (d -2) \Gamma \left(\frac{3 d }{2}-3\right)}\,, \\
M_{3,3}
& = \
\int_{\mathbf{q}_1,\mathbf{q}_2,\mathbf{q}_3}  \frac{1}{\mathbf{q}_1^2\mathbf{q}_3^2\left(\mathbf{q}_1-\mathbf{q}_2\right)^2\left(\mathbf{q}_2+\mathbf{p}\right)^2\left(\mathbf{q}_3-\mathbf{p}\right)^2}
= \frac{|\mathbf{p}|^{-10+3d}}{\left( 4 \pi \right)^{\frac{3 d}{2}}} \frac{\Gamma \left(5-\frac{3 d }{2}\right) \Gamma \left(2-\frac{d }{2}\right)^2
   \Gamma \left(\frac{d }{2}-1\right)^5 \Gamma \left(\frac{3 d }{2}-4\right)}{\Gamma (4-d ) \Gamma (d -2)^2 \Gamma (2 d -5)}\,,
\end{align}
\end{subequations}

where the final integration over the total exchanged momenta, $\mathbf{p}$, results in a Fourier transform producing the expected $1/r^n$-type potential. To obtain the final result we must also include corrections to the kinetic term,
\begin{equation}
  K(t)=\sum_{a=1,2} m_a \left(\frac{1}{2} \mathbf{v}_a^2+\frac{1}{8} \mathbf{v}_a^4+\frac{1}{16} \mathbf{v}_a^6+ \frac{5}{128} \mathbf{v}_a^8+\mathcal{O}\left(\mathbf{v}^{10}\right)\right)\,.
\end{equation}
As it is well known, the potential features a divergent term, entering as a pole in the $d\to 3$ limit \cite{Foffa:2011ub}. Following \cite{Goldberger:2004jt,Goldberger:2009qd}, we introduce an arbitrary regularization scale $\mu$, associated with a $d$ dimensional Newton's constant $G \to \mu^{d-3}G$. We display below only the part stemming off of the $1/(d-3)$ pole,
\begin{align}\label{eq:poled_lagr}
\mathcal{L}^{\text{(pole)}}_{3\rm PN}
= & \nm
\frac{G^4 m_1 m_2}{3 r^4}[m_1^3+m_2^3+7 m_1 m_2(m_1+m_2)]\left(\frac{1}{d-3}-2\Log{\left(\mu^2_s r^2\right)}\right)\\
 & \nm 
+\frac{G^3 m_1 m_2}{18 r^3} \left(\frac{2}{d-3}-3\Log{\left(\mu^2_s r^2\right)}\right)\bigl\{[21 m_1^2+24m_1 m_2+5m_2^2]\, r\left(\mathbf{n} \cdot \mathbf{a}_1 \right)-2 m_2[12 m_1+13 m_2 ]\, r\left(\mathbf{n} \cdot \mathbf{a}_2 \right)\\
 & 
+[3 m_1^2-24 m_1 m_2 -2m_2^2]\bigl(3 \left(\mathbf{n} \cdot \mathbf{v}_1 \right)^2+3 \left(\mathbf{n} \cdot \mathbf{v}_2 \right)^2- \mathbf{v}_1^2-\mathbf{v}_2^2-6\left(\mathbf{n} \cdot \mathbf{v}_1 \right) \left(\mathbf{n} \cdot \mathbf{v}_2\right)+2\left(\mathbf{v}_1 \cdot \mathbf{v}_2 \right)\bigr) \bigr\}\,,
\end{align}
where $\mu_s \equiv  \sqrt{4 \pi} e^{\frac{\gamma_E}{2}} \mu$ \cite{Foffa:2019yfl}, with $\gamma_E$ being the Euler-Mascheroni constant.
 The expression for the full Lagrangian is given in the ancillary file, together with the equations of motion for both compact objects, as well as in the center-of-mass.\vskip 4pt Similarly to the procedure outlined in \cite{Foffa:2011ub,Blanchet:2013haa}, both the divergence and logarithm can be removed by a coordinate shift, in our case,
\begin{subequations}\label{eq:3pn_shift}
\begin{align}
&
\mathbf{x}^i_1 \rightarrow \mathbf{x}^i_1 -\frac{G^3 m_2 (7m_1^2+m_2^2)}{6 r^3}\left(-\frac{2}{d-3}+3 \Log{\left(\mu_s^2 r^2\right)} \right)\mathbf{r}^i-\frac{G^2 m_1^2}{6}\left(-\frac{2}{d-3}+2\Log{\left(\mu_s^2 r^2\right)} \right)\mathbf{a}_1^i\,,\\
&
\mathbf{x}^i_2 \rightarrow \mathbf{x}^i_2 +\frac{G^3 m_1 (7m_2^2+m_1^2)}{6 r^3}\left(-\frac{2}{d-3}+3 \Log{\left(\mu_s^2 r^2\right)} \right)\mathbf{r}^i-\frac{G^2 m_2^2}{6}\left(-\frac{2}{d-3}+2\Log{\left(\mu_s^2 r^2\right)} \right)\mathbf{a}_2^i\,,
\end{align}
\end{subequations}
implemented into the leading order Lagrangian, yielding a finite result. For instance, the equations of motion in the center-of-mass frame take the form
\begin{align}\label{eq:3pn_eom}
\mathbf{a}^i_{3\rm PN}\left(\mathbf{r},\mathbf{v}\right)
= & \nm
\biggl\{ \frac{G^4 M^4 \nu}{16\, r^6}\left(2164-41 \pi ^2+568 \nu\right)\\
 & \nm
-\frac{G^3 M^3}{192 \, r^5} \bigl[\left( -9024+5 \left(-248+81 \pi ^2\right) \nu +264 \nu^2-1344 \nu^3 \right)\left(\mathbf{n} \cdot \mathbf{v} \right)^2 \\
& \nm
+\left(1920+\left(5864-81 \pi ^2\right) \nu +192 \nu ^3\right) \mathbf{v}^2\bigr]\\
& \nm
+\frac{G^2 M^2 \nu}{4 \, r^4} \bigl[ 2 \left(2+69 \nu +60 \nu^2\right) \left(\mathbf{n} \cdot \mathbf{v} \right)^4 -4 \left(15+16 \nu +20 \nu^2\right)\left(\mathbf{n} \cdot \mathbf{v} \right)^2\mathbf{v}^2 \\
& \nm
+\left(21-32 \nu +40 \nu^2\right)\mathbf{v}^4 \bigr]\\
& \nm
+\frac{G M \nu}{16 \, r^3} \bigl[ 35 \left(1-5 \nu +5 \nu ^2\right) \left(\mathbf{n} \cdot \mathbf{v} \right)^6 -30 \left(4-18 \nu +17 \nu ^2\right)\left(\mathbf{n} \cdot \mathbf{v} \right)^4 \mathbf{v}^2 \\
& \nm
+6 \left(20-79 \nu +60 \nu ^2\right) \left(\mathbf{n} \cdot \mathbf{v} \right)^2 \mathbf{v}^4 -4 \left(11-49 \nu +52 \nu ^2\right) \mathbf{v}^6 \bigr]\biggr\}\,\mathbf{r}^i\\
& \nm
+\biggl\{ \frac{G^3 M^3}{96 \, r^4} \left(-3456+\left(8092+81 \pi^2\right) \nu -2400 \nu^2-768 \nu ^3 \right)\left(\mathbf{n} \cdot \mathbf{v} \right) \\
& \nm
+\frac{G^2 M^2 \nu}{6 \, r^3} \bigl[ \left(407-177 \nu -108 \nu^2\right) \left(\mathbf{n} \cdot \mathbf{v} \right)^3 +6 \left(-49+27 \nu +10 \nu ^2\right)\left(\mathbf{n} \cdot \mathbf{v} \right)\mathbf{v}^2\bigr]\\
& \nm
-\frac{G M \nu}{8 \, r^2} \bigl[ 15 \left(-3+8 \nu +2 \nu ^2\right) \left(\mathbf{n} \cdot \mathbf{v} \right)^5 -6 \left(-16+37 \nu +16 \nu ^2\right)\left(\mathbf{n} \cdot \mathbf{v} \right)^3 \mathbf{v}^2\bigr]\\
&
+\left(-65+152 \nu +48 \nu ^2\right) \left(\mathbf{n} \cdot \mathbf{v} \right) \mathbf{v}^4 \biggr\} \mathbf{v}^i \,,
\end{align}
with lower order equations of motion substitutions and center-of-mass corrections. The above acceleration can be shown to be equivalent (upon coordinate transformations) to the known expressions in \cite{Foffa:2011ub,Blanchet:2013haa}.\vskip 4pt 

It is straightforward to derive the corresponding EFT Hamiltonian, via a Legendre transformation 
\begin{align}\label{eq:ham_formula}
\mathcal{H}
= 
& \nm
\sum_{a=1,2} v_a^i \left(\frac{\partial \mathcal{L}}{\partial v_a^i}\right)+a_a^i \left(\frac{\partial \mathcal{L}}{\partial a_a^i}\right)+\dot{a}_a^i \left(\frac{\partial \mathcal{L}}{\partial \dot{a}_a^i}\right)-v_a^i \left(\partial_t \frac{\partial \mathcal{L}}{\partial a_a^i} \right)-a_a^i \left(\partial_t \frac{\partial \mathcal{L}}{\partial \dot{a}_a^i} \right)+v_a^i \left(\partial^2_t \frac{\partial \mathcal{L}}{\partial \dot{a}_a^i} \right)\\
&
-\mathcal{L}\left(\mathbf{r},\mathbf{v}_1,\mathbf{a}_1,\dot{\mathbf{a}}_1,\mathbf{v}_2,\mathbf{a}_2,\dot{\mathbf{a}}_2\right)\,.
\end{align}
We limit ourselves here to the center-of-mass frame, resulting in 
\begin{align}\label{eq:3pn_hamiltonian}
\mathcal{H}_{3 \rm PN}\left(\mathbf{r},\mathbf{v}\right)
= & \nm
-\frac{G^4 M^5 \nu}{72\, r^4}\left[\left(36 \pi^2-991\right) \nu +309\right]\\
 & \nm
+\frac{G^3 M^4 \nu}{576 \, r^3} \bigl[3 \left(-480+\left(2296-81 \pi^2\right) \nu +2448 \nu^2+672 \nu^3\right)\left(\mathbf{n} \cdot \mathbf{v} \right)^2 \\
& \nm
+\left(912+\left(-2296+81 \pi^2\right) \nu -3024 \nu^2+288 \nu^3\right)\mathbf{v}^2\bigr]\\
& \nm
+\frac{G^2 M^3 \nu}{48 \, r^2} \bigl[ \nu  \left(-91+492 \nu +288 \nu^2\right) \left(\mathbf{n} \cdot \mathbf{v} \right)^4 -3 \left(84-696 \nu +815 \nu ^2+324 \nu ^3\right)\left(\mathbf{n} \cdot \mathbf{v} \right)^2\mathbf{v}^2 \\
& \nm
+3 \left(183-498 \nu +406 \nu ^2-108 \nu ^3\right) \mathbf{v}^4 \bigr]\\
& \nm
+\frac{G M^2 \nu}{16 \, r} \bigl[ 5 \nu  \left(1-5 \nu +5 \nu^2\right) \left(\mathbf{n} \cdot \mathbf{v} \right)^6 -3 \nu  \left(3-28 \nu +55 \nu ^2\right)\left(\mathbf{n} \cdot \mathbf{v} \right)^4 \mathbf{v}^2 \\
& \nm
+3 \nu  \left(-7-25 \nu +125 \nu ^2\right) \left(\mathbf{n} \cdot \mathbf{v} \right)^2 \mathbf{v}^4 +\left(55-215 \nu +116 \nu ^2+325 \nu ^3\right) \mathbf{v}^6 \bigr]\\
&
-\frac{7 M \nu}{128} \left( -5+59 \nu -238 \nu ^2+323 \nu ^3\right) \mathbf{v}^8\,.
\end{align}
From here we obtain the (gauge-invariant) binding energy for quasi-circular orbits to 3PN order,
\begin{align}\label{eq:3pnham_circ}
E_{\rm circ}
= & \nm
-\frac{M \nu x}{2}\,\bigg\{1+\left(-\frac{3}{4}-\frac{\nu }{12} \right)x +\left( -\frac{27}{8}+\frac{19}{8}\nu -\frac{\nu ^2}{24}\right)x^2\\
 & 
 +\left[-\frac{675}{64}+\left(\frac{34445}{576}-\frac{205 \pi ^2 }{96}\right)\nu-\frac{155}{96}\nu^2-\frac{35}{5184}\nu^3 \right]x^3 + {\cal O}(x^4)\bigg\}\,,
\end{align}
in terms of the PN parameter $x\equiv\left(G M \omega \right)^{2/3}$, with $\omega$ the orbital frequency. As expected, this result agrees with the value given in \cite{Blanchet:2004ek,Foffa:2011ub,Schafer:2018jfw}.

\section{Radiative dynamics at N$^3$LO}\label{sec:rad}

We now move onto the novel aspects of the work and the derivation of gravitational radiation at 3PN order. As shown in \cite{Goldberger:2009qd}, UV divergences emerge in the derivation of the ``tail-of-tail'' effect, which anticipate the existence of similar UV poles in the matching of the quadrupole moment. This is expected, since  the UV pole must cancel out in a {\it full theory} computation.\footnote{IR divergences are also present, but these either cancel in the derivation of the final observable or can be reabsorbed into a phase/time redefinition, e.g. \cite{Porto:2012as}.} Because of this, it is vital to conduct a multipole expansion of the effective action in $d$ dimensions, so that moments of the stress-energy tensor must be decomposed into irreducible tensors under SO($d$). As shown in \cite{Amalberti:2023ohj}, the key contribution to the mass quadrupole moment at 3PN order becomes
\begin{align}\label{eq:3pn_mq_symbolic}
I^{ij}_{3\rm PN}
= & \nm
\left[\int\!\! \dd^3 \mathbf{x}\, T^{00}_{3\rm PN}\mathbf{x}^i\mathbf{x}^j\right]_{\text{TF}}+\left(1-(d-3)\right)\left[\int\!\! \dd^3 \mathbf{x}\, T^{a a}_{2\rm PN}\mathbf{x}^i\mathbf{x}^j\right]_{\text{TF}} - \left(\frac{4}{3}-\frac{10}{9}\,(d-3)\right)\left[\int\!\! \dd^3 \mathbf{x}\, \partial_t\tilde{T}^{0}_{2\rm PN}\mathbf{x}^i\mathbf{x}^j\right]_{\text{TF}}\\
& \nm
+\left(\frac{1}{6}-\frac{13}{72}\,(d-3) \right)\left[\int\!\! \dd^3 \mathbf{x}\, \partial_t^2\tilde{T}_{1\rm PN}\mathbf{x}^i\mathbf{x}^j\right]_{\text{TF}}+\left(\frac{11}{42}-\frac{173}{882}\,(d-3) \right)\left[\int\!\! \dd^3 \mathbf{x}\, \partial_t^2T^{00}_{2\rm PN} r^2 \mathbf{x}^i\mathbf{x}^j\right]_{\text{TF}}\\
& \nm
+\left(\frac{2}{21}-\frac{391}{3528}\,(d-3) \right)\left[\int\!\! \dd^3 \mathbf{x}\, \partial_t^2 T^{kk}_{1\rm PN} r^2 \mathbf{x}^i\mathbf{x}^j\right]_{\text{TF}} - \frac{1}{7}\left[\int\!\! \dd^3 \mathbf{x}\, \partial_t^3 \tilde{T}^{0}_{1\rm PN} r^2 \mathbf{x}^i\mathbf{x}^j\right]_{\text{TF}}\\
& \nm
+ \frac{1}{84}\left[\int\!\! \dd^3 \mathbf{x}\, \partial_t^4 \tilde{T}_{0\rm PN} r^2 \mathbf{x}^i\mathbf{x}^j\right]_{\text{TF}} + \frac{23}{1512}\left[\int\!\! \dd^3 \mathbf{x}\, \partial_t^4 T^{00}_{1\rm PN} r^4 \mathbf{x}^i\mathbf{x}^j\right]_{\text{TF}}+ \frac{5}{1512}\left[\int\!\! \dd^3 \mathbf{x}\, \partial_t^4 T^{a a}_{0\rm PN} r^4 \mathbf{x}^i\mathbf{x}^j\right]_{\text{TF}}\\
& \nm
- \frac{1}{189}\left[\int\!\! \dd^3 \mathbf{x}\, \partial_t^5 \tilde{T}^0_{0PN} r^4 \mathbf{x}^i\mathbf{x}^j\right]_{\text{TF}} + \frac{13}{33264}\left[\int\!\! \dd^3 \mathbf{x}\, \partial_t^6 T^{00}_{0\rm PN} r^6 \mathbf{x}^i\mathbf{x}^j\right]_{\text{TF}} \\ 
&
+ \text{lower order corrections}\,, 
\end{align}
where $T_{n\rm PN}^{ab}$ corresponds to the $n$PN order in the matching of the stress-energy tensor and, for simplicity, we have used the shorthanded notation $\tilde{T}^0 \equiv T^{0k}x^k$, $\tilde{T} \equiv T^{k\ell}x^kx^\ell$. The expressions for the lower order corrections to the quadrupole moments, $I^{ij}_{1\rm PN}$ and $I^{ij}_{2\rm PN}$, in terms of moments of the stress-energy tensor can be found in \cite{Goldberger:2009qd,Leibovich:2019cxo}, and must be evaluated using the equations of motion up to the 2PN order, starting from $a_{\rm N}$ (the Newtonian acceleration), which must be also evaluated in $d$ dimensions,
\begin{equation}\label{eq:shift_0pneom}
    \mathbf{a}^i_{\rm N}\left(\mathbf{r}\right)=-\frac{G M}{r^3}\left\{1+\frac{d-3}{2}\left[3-\Log{\left( \mu_s^2 r^2\right)} \right]\right\}\mathbf{r}^i\,.
\end{equation}

\begin{figure}[t!]
  \centering
  \begin{subfigure}[b]{0.13\textwidth}
    \includegraphics[width=\linewidth]{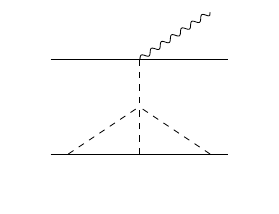}
  \end{subfigure}
  \begin{subfigure}[b]{0.13\textwidth}
    \includegraphics[width=\linewidth]{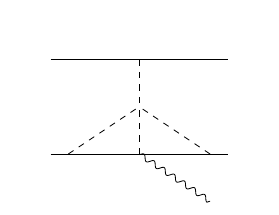}
  \end{subfigure}
  \begin{subfigure}[b]{0.13\textwidth}
    \includegraphics[width=\linewidth]{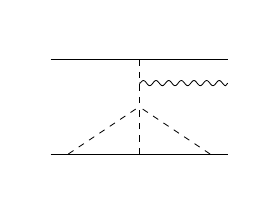}
  \end{subfigure}
  \begin{subfigure}[b]{0.13\textwidth}
    \includegraphics[width=\linewidth]{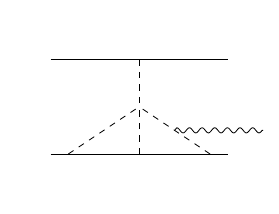}
  \end{subfigure}
  \begin{subfigure}[b]{0.13\textwidth}
    \includegraphics[width=\linewidth]{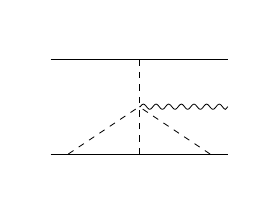}
  \end{subfigure}
    \begin{subfigure}[b]{0.13\textwidth}
    \includegraphics[width=\linewidth]{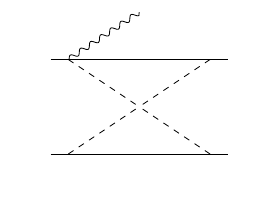}
  \end{subfigure}
  \begin{subfigure}[b]{0.13\textwidth}
    \includegraphics[width=\linewidth]{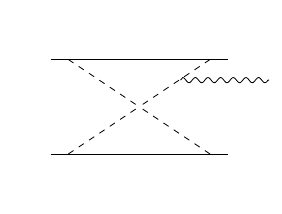}
  \end{subfigure}
  \begin{subfigure}[b]{0.13\textwidth}
    \includegraphics[width=\linewidth]{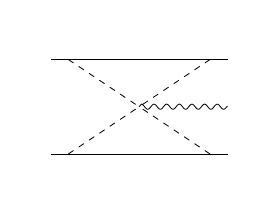}
  \end{subfigure}
  \begin{subfigure}[b]{0.13\textwidth}
    \includegraphics[width=\linewidth]{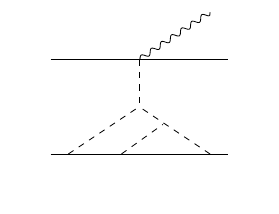}
  \end{subfigure}
  \begin{subfigure}[b]{0.13\textwidth}
    \includegraphics[width=\linewidth]{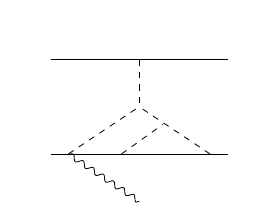}
  \end{subfigure}
    \begin{subfigure}[b]{0.13\textwidth}
    \includegraphics[width=\linewidth]{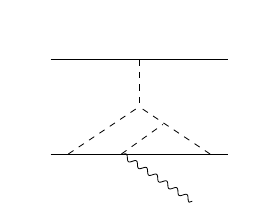}
  \end{subfigure}
  \begin{subfigure}[b]{0.13\textwidth}
    \includegraphics[width=\linewidth]{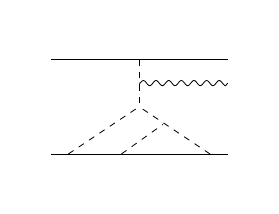}
  \end{subfigure}
  \begin{subfigure}[b]{0.13\textwidth}
    \includegraphics[width=\linewidth]{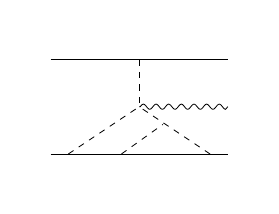}
  \end{subfigure}
  \begin{subfigure}[b]{0.13\textwidth}
    \includegraphics[width=\linewidth]{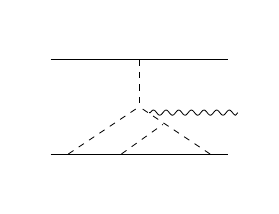}
  \end{subfigure}
  \begin{subfigure}[b]{0.13\textwidth}
    \includegraphics[width=\linewidth]{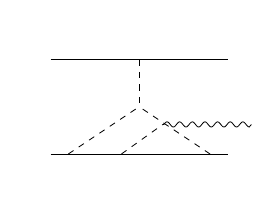}
  \end{subfigure}
    \begin{subfigure}[b]{0.13\textwidth}
    \includegraphics[width=\linewidth]{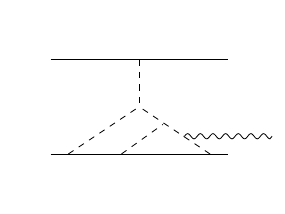}
  \end{subfigure}
  \begin{subfigure}[b]{0.13\textwidth}
    \includegraphics[width=\linewidth]{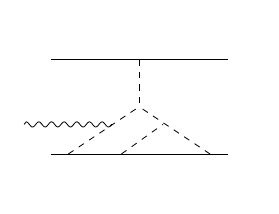}
  \end{subfigure}
  \begin{subfigure}[b]{0.13\textwidth}
    \includegraphics[width=\linewidth]{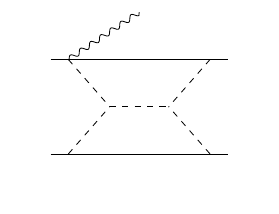}
  \end{subfigure}
  \begin{subfigure}[b]{0.13\textwidth}
    \includegraphics[width=\linewidth]{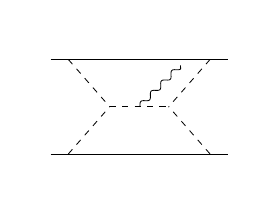}
  \end{subfigure}
  \begin{subfigure}[b]{0.13\textwidth}
    \includegraphics[width=\linewidth]{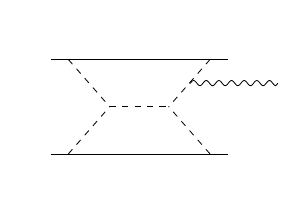}
  \end{subfigure}
    \begin{subfigure}[b]{0.13\textwidth}
    \includegraphics[width=\linewidth]{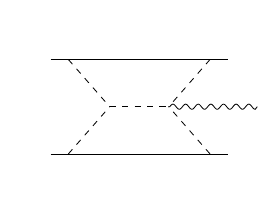}
  \end{subfigure}
  \begin{subfigure}[b]{0.13\textwidth}
    \includegraphics[width=\linewidth]{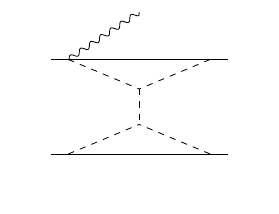}
  \end{subfigure}
  \begin{subfigure}[b]{0.13\textwidth}
    \includegraphics[width=\linewidth]{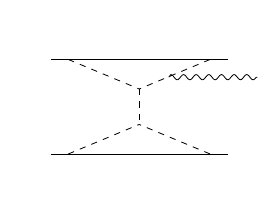}
  \end{subfigure}
  \begin{subfigure}[b]{0.13\textwidth}
    \includegraphics[width=\linewidth]{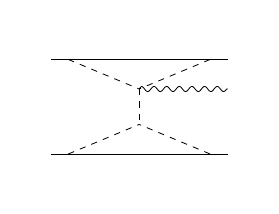}
  \end{subfigure}
  \begin{subfigure}[b]{0.13\textwidth}
    \includegraphics[width=\linewidth]{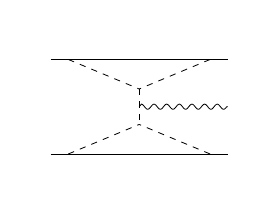}
  \end{subfigure}
    \begin{subfigure}[b]{0.13\textwidth}
    \includegraphics[width=\linewidth]{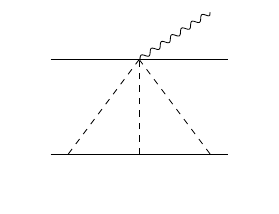}
  \end{subfigure}
  \begin{subfigure}[b]{0.13\textwidth}
    \includegraphics[width=\linewidth]{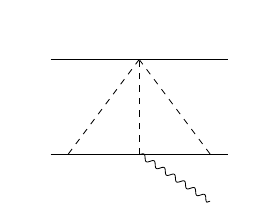}
  \end{subfigure}
  \begin{subfigure}[b]{0.13\textwidth}
    \includegraphics[width=\linewidth]{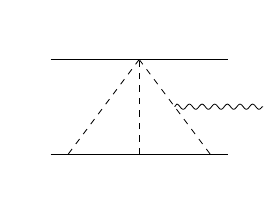}
  \end{subfigure}
  \begin{subfigure}[b]{0.13\textwidth}
    \includegraphics[width=\linewidth]{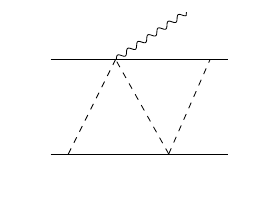}
  \end{subfigure}
  \begin{subfigure}[b]{0.13\textwidth}
    \includegraphics[width=\linewidth]{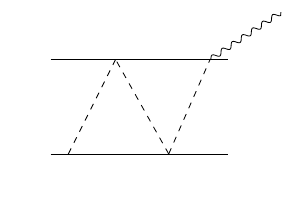}
  \end{subfigure}
    \begin{subfigure}[b]{0.13\textwidth}
    \includegraphics[width=\linewidth]{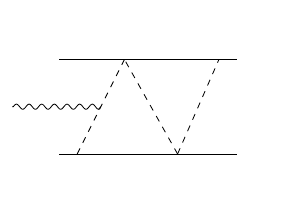}
  \end{subfigure}
  \begin{subfigure}[b]{0.13\textwidth}
    \includegraphics[width=\linewidth]{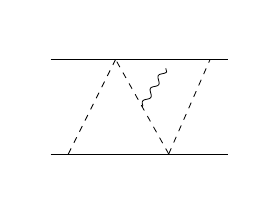}
  \end{subfigure}
  \begin{subfigure}[b]{0.13\textwidth}
    \includegraphics[width=\linewidth]{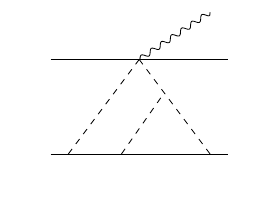}
  \end{subfigure}
  \begin{subfigure}[b]{0.13\textwidth}
    \includegraphics[width=\linewidth]{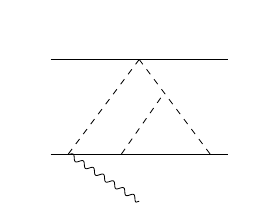}
  \end{subfigure}
  \begin{subfigure}[b]{0.13\textwidth}
    \includegraphics[width=\linewidth]{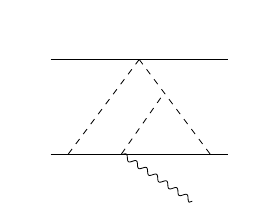}
  \end{subfigure}
    \begin{subfigure}[b]{0.13\textwidth}
    \includegraphics[width=\linewidth]{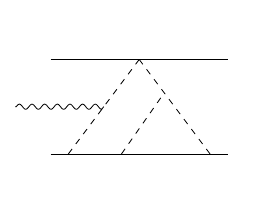}
  \end{subfigure}
  \begin{subfigure}[b]{0.13\textwidth}
    \includegraphics[width=\linewidth]{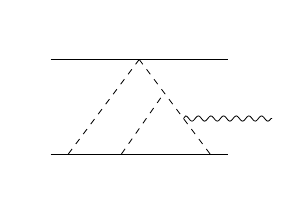}
  \end{subfigure}
  \begin{subfigure}[b]{0.13\textwidth}
    \includegraphics[width=\linewidth]{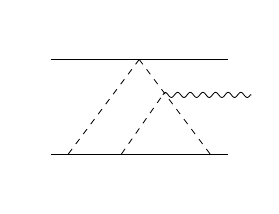}
  \end{subfigure}
  \begin{subfigure}[b]{0.13\textwidth}
    \includegraphics[width=\linewidth]{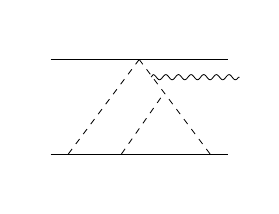}
  \end{subfigure}
  \begin{subfigure}[b]{0.13\textwidth}
    \includegraphics[width=\linewidth]{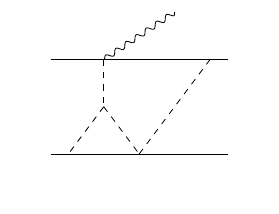}
  \end{subfigure}
    \begin{subfigure}[b]{0.13\textwidth}
    \includegraphics[width=\linewidth]{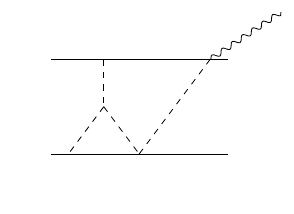}
  \end{subfigure}
  \begin{subfigure}[b]{0.13\textwidth}
    \includegraphics[width=\linewidth]{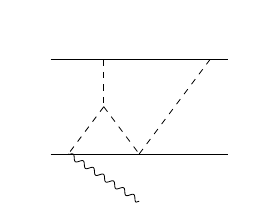}
  \end{subfigure}
  \begin{subfigure}[b]{0.13\textwidth}
    \includegraphics[width=\linewidth]{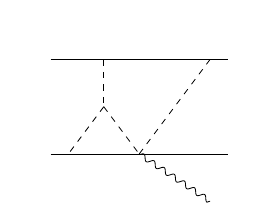}
  \end{subfigure}
  \begin{subfigure}[b]{0.13\textwidth}
    \includegraphics[width=\linewidth]{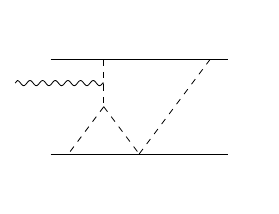}
  \end{subfigure}
    \begin{subfigure}[b]{0.13\textwidth}
    \includegraphics[width=\linewidth]{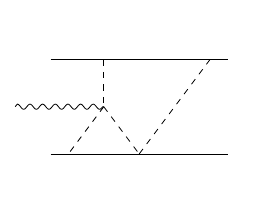}
  \end{subfigure}
  \begin{subfigure}[b]{0.13\textwidth}
    \includegraphics[width=\linewidth]{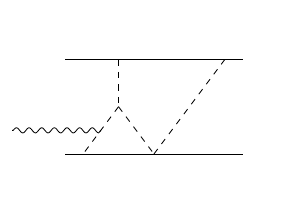}
  \end{subfigure}
  \begin{subfigure}[b]{0.13\textwidth}
    \includegraphics[width=\linewidth]{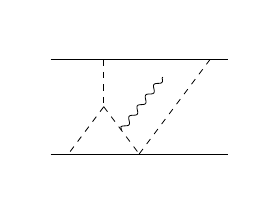}
  \end{subfigure}
  \begin{subfigure}[b]{0.13\textwidth}
    \includegraphics[width=\linewidth]{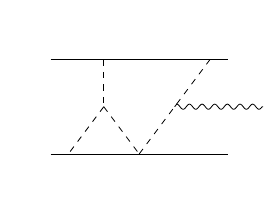}
  \end{subfigure} 
  \caption{Feynman topologies relevant for the matching of the stress-energy tensor. The wavy lines represents the radiation field, see also e.g. \cite{Goldberger:2004jt,Goldberger:2009qd,Porto:2016pyg,Leibovich:2019cxo} (mirror images are omitted).}
  \label{fig:radiation_3pn}
\end{figure}

In order to obtain the moments needed in~\eqref{eq:3pn_mq_symbolic} it is useful to introduce the mixed Fourier transform, ${T}^{\mu \nu}\left(t, \mathbf{k}\right)$ introduced in \cite{Goldberger:2009qd}, and expand in the long-wavelength approximation, in powers of  $(\mathbf{k}\cdot \mathbf{x}) \sim \mathbf{v} \ll 1$, such that
\begin{equation}\label{eq:matching_stress}
    {T}^{\mu \nu}\left(t, \mathbf{k}\right)=\sum^{\infty}_{n=0}\,\frac{(-i)^n}{n!}\left(\int \, \dd^3\mathbf{x} \,T^{\mu \nu}(t,\mathbf{x}) \mathbf{x}^{i_1}\mathbf{x}^{i_2}\dots \mathbf{x}^{i_n}\right)\mathbf{k}^{i_1}\mathbf{k}^{i_2}\dots \mathbf{k}^{i_n}\,,
\end{equation}
The topologies needed to obtain the stress-energy tensor, beyond those discussed in \cite{Leibovich:2019cxo}, are shown in Fig.~\ref{fig:radiation_3pn}. The list of vertices expressing the couplings between worldlines, potential and radiation modes is detailed in Appendix~\ref{app:sub_rad}.\vskip 4pt

As discussed in \cite{Goldberger:2004jt,Goldberger:2009qd}, in the matching computation we encounter couplings between potential and radiation modes, in which the momenta of some of the potential line(s) does not scale homogeneously with the velocity. To solve this problem, it is then customary to expand the propagator in terms of the radiation momentum, as 
\begin{equation}\label{eq:external_denom}
    \frac{1}{\left(\mathbf{q}+\mathbf{k}\right)^2}=\frac{1}{\mathbf{q}^2}-\frac{2\, \mathbf{q} \cdot \mathbf{k}}{\mathbf{q}^4}-\frac{\mathbf{k}^2}{\mathbf{q}^4}+\frac{4\left(\mathbf{q} \cdot \mathbf{k}\right)^3}{\mathbf{q}^6}+\frac{4\,\mathbf{k}^2\left(\mathbf{q}\cdot \mathbf{k}\right)}{\mathbf{q}^6}-\frac{8 \left(\mathbf{q}\cdot \mathbf{k}\right)^3}{\mathbf{q}^8}+\cdots\,.
\end{equation}
It is worth noticing that that the expansion in~\eqref{eq:external_denom} mixes the {\it internal} (potential) and {\it external} (radiation) momenta. As a result, the construction of the integrand often requires a tensor reduction. Similarly to the conservative case, all the resulting  (two-loop) Feynman scalar integrals can be reduced to master integrals by employing IBPs relations. In particular, for the diagrams in Fig.~\ref{fig:radiation_3pn} we simply need,
\begin{subequations}\label{eq:2loop_mi}
\begin{align}
M_{2,1}
& = 
\int_{\mathbf{q}_1,\mathbf{q}_2} \frac{1}{\mathbf{q}_1^2\mathbf{q}_2^2\left(\mathbf{q}_1+\mathbf{q}_2-\mathbf{p}\right)^2}= \frac{|\mathbf{p}|^{-6+2d}}{\left( 4 \pi \right)^{d}} \frac{\Gamma (3-d ) \Gamma \left(\frac{d }{2}-1\right)^3 }{\Gamma \left(\frac{3 d }{2}-3\right)} \,, \\
M_{2,2}
& = 
\int_{\mathbf{q}_1,\mathbf{q}_2} \frac{1}{\mathbf{q}_1^2\left(\mathbf{q}_1+\mathbf{q}_2\right)^2\left(\mathbf{q}_1-\mathbf{p}\right)^2\left(\mathbf{q}_1+\mathbf{q}_2-\mathbf{p}\right)^2}= \frac{|\mathbf{p}|^{-8+2d}}{\left( 4 \pi \right)^{d}} \frac{\Gamma \left(2-\frac{d }{2}\right)^2 \Gamma \left(\frac{d }{2}-1\right)^4}{\Gamma (d -2)^2} \,,
\end{align}
\end{subequations}
The computation of the remaining topologies is discussed in  Appendix~\ref{app:sub_rad}.\vskip 4pt

Gathering all the pieces together, we obtain the contribution to the quadrupole moment to 3PN order. However, our result still depends on the original (harmonic) coordinate system. In order to be compatible with the (UV finite) result for the equations of motion from the conservative sector, we must also incorporate the coordinate shift in~\eqref{eq:3pn_shift}, which enters in the leading quadrupole moment as 
\begin{equation}\label{eq:3pn_mq_shift}
I^{ij}_{\rm LO}
\rightarrow
M \nu \,\mathbf{r}^i\mathbf{r}^j +\frac{G^3 M^4 \nu}{r^3}\left\{-\nu+\left(1+3\nu\right)\left[\frac{2}{3(d-3)}-\Log{\left(\mu_s^2 r^2\right)}\right]\right\}\,\mathbf{r}^i\mathbf{r}^j\,,
\end{equation}
yielding for the quadrupole moment at 3PN the final result:
\begin{align}\label{eq:3pn_mq}
I^{ij}_{\rm 3PN}\left(\mathbf{r},\mathbf{v}\right)
= & \nm
\biggl\{\biggl\{\frac{G^3 M^4 \nu}{1455300\, r^3}\left[\left(6093623-1050 \left(-36181+1386 \pi ^2\right) \nu -7627200 \nu ^2+2325575 \nu ^3\right)\right]\\
 & \nm
-\frac{G^2 M^3 \nu}{83160 \, r^2} \bigl[\left(303436-1296500 \nu +1674720 \nu ^2+811815 \nu ^3\right)\left(\mathbf{n} \cdot \mathbf{v} \right)^2 \\
& \nm
-\left(646543+2963735 \nu +2174545 \nu ^2-1121070 \nu ^3\right)\mathbf{v}^2\bigr]\\
& \nm
+\frac{G M^2 \nu}{16632 \, r} \bigl[ \left(336-10470 \nu +48957 \nu ^2-52248 \nu ^3\right) \left(\mathbf{n} \cdot \mathbf{v} \right)^4 \\
& \nm
-\left(2271+11090 \nu +98311 \nu ^2-306814 \nu ^3\right)\left(\mathbf{n} \cdot \mathbf{v} \right)^2\mathbf{v}^2 \\
& \nm
+2 \left(33156-188950 \nu +218411 \nu ^2+299857 \nu ^3\right) \mathbf{v}^4 \bigr]\\
& \nm
+\frac{M \nu}{11088} \left(4561-55951 \nu +234134 \nu ^2-328663 \nu ^3\right)\mathbf{v}^6\biggr\}\mathbf{r}^i\mathbf{r}^j\\
& \nm
+\biggl\{\frac{G^2 M^3 \nu}{363825} \left(1957143-6457150 \nu +10950800 \nu ^2-4254950 \nu ^3\right) \\
& \nm
+\frac{G M^2 \nu \,r}{4158} \bigl[ 3 \left(1422-5807 \nu +515 \nu ^2+16490 \nu ^3\right) \left(\mathbf{n} \cdot \mathbf{v} \right)^2 \\
& \nm
+\left(15849-67933 \nu +51320 \nu ^2+129781 \nu ^3\right)\mathbf{v}^2 \bigr]\\
& \nm
+\frac{M \nu \, r^2}{5544} \bigl[ 20 \left(23-227 \nu +718 \nu ^2-689 \nu ^3\right)\left(\mathbf{n} \cdot \mathbf{v} \right)^2\mathbf{v}^2\\
& \nm
+\left(1369-19351 \nu +90842 \nu ^2-139999 \nu ^3\right)\mathbf{v}^4\bigr]\biggr\}\mathbf{v}^i\mathbf{v}^j\\
& \nm
+\biggl\{\frac{G^2 M^3 \nu}{41580 \, r} \left(-294454+1287185 \nu -569730 \nu ^2+133230 \nu ^3\right)\left(\mathbf{n} \cdot \mathbf{v} \right) \\
& \nm
-\frac{G M^2 \nu \,r}{8316} \bigl[ 3 \left(-305-3233 \nu +8611 \nu ^2+32220 \nu ^3\right) \left(\mathbf{n} \cdot \mathbf{v} \right)^3 \\
& \nm
+\left(34068-237893 \nu +376126 \nu ^2+234260 \nu ^3\right)\left(\mathbf{n} \cdot \mathbf{v} \right)\mathbf{v}^2 \bigr]\\
& \nm
+\frac{M \nu \, r }{1386} \left(-457+6103 \nu -27386 \nu ^2+40687 \nu ^3\right)\left(\mathbf{n} \cdot \mathbf{v} \right)\mathbf{v}^4\biggr\}\mathbf{r}^i\mathbf{v}^j\\
&
+\frac{107\,G^3 M^4 \nu}{105\, r^3}\left(\frac{2}{d-3} -3\Log{\left( \mu_s^2 r^2\right)} \right)\mathbf{r}^i\mathbf{r}^j-\frac{214\,G^2 M^3 \nu}{105\, r^3}\left(\frac{1}{d-3} -\Log{\left( \mu_s^2 r^2\right)} \right)\mathbf{v}^i\mathbf{v}^j\biggr\}_{\text{STF}}\,,
\end{align}
 Armed with the source multipole moments\footnote{The computation of the other necessary multipoles (needed at lower PN order) is discussed in App.~\ref{appb3}.} we can readily derive the contribution to the GW flux via, e.g. ~\cite{Porto:2016pyg},
\begin{align}\label{eq:3pn_flux_symbolic}
\mathcal{P}^{(\text{inst})}
= & \nm
\,G \bigg\{ \frac{1}{5}I_{ij}^{(3)}I_{ij}^{(3)}+\frac{1}{189}I_{ijk}^{(4)}I_{ijk}^{(4)}+\frac{1}{9072}I_{ijkl}^{(5)}I_{ijkl}^{(5)}+\frac{1}{594000}I_{ijklm}^{(6)}I_{ijklm}^{(6)}\\
&
+\frac{16}{45}J_{ij}^{(3)}J_{ij}^{(3)}+\frac{1}{84}J_{ijk}^{(4)}J_{ijk}^{(4)}+\frac{4}{14175}J_{ijkl}^{(5)}J_{ijkl}^{(5)}+\dots\bigg\}\,,
\end{align}
corresponding to {\it instantaneous} (source) terms. The expression at 3PN, evaluated in the center-of-mass, is given by 
\begin{align}\label{eq:3pn_flux}
\mathcal{P}^{(\text{inst})}_{3\rm PN}\left(\mathbf{r},\mathbf{v}\right)
= & \nm
-\frac{G^7 M^8 \nu^2}{31185\, r^8}\left[-98457+361912 \nu +126798 \nu ^2+3464 \nu ^3\right]\\
 & \nm
+\frac{G^6 M^7 \nu^2}{10914750 \, r^7} \bigl[\left(4599453072-1925 \left(-8448536+357399 \pi ^2\right) \nu +2176410600 \nu ^2-481801600 \nu ^3\right))\left(\mathbf{n} \cdot \mathbf{v} \right)^2 \\
& \nm
+\left(-5704087808+9625 \left(-1664440+58401 \pi ^2\right) \nu +563220000 \nu ^2-25902800 \nu ^3\right)\mathbf{v}^2\bigr]\\
& \nm
+\frac{G^5 M^6 \nu^2}{10914750 \, r^6} \bigl[ \left(-210 \left(-134574346+15 \left(2191804+483021 \pi ^2\right) \nu -76954850 \nu ^2+16506620 \nu ^3\right)\right) \left(\mathbf{n} \cdot \mathbf{v} \right)^4\\
& \nm
+6 \left(-6326624876+1575 \left(652462+198891 \pi ^2\right) \nu -2855456100 \nu ^2+694513400 \nu ^3\right)\left(\mathbf{n} \cdot \mathbf{v} \right)^2\mathbf{v}^2 \\
& \nm
+\left(10245313964-525 \left(4720384+767151 \pi ^2\right) \nu +2731904700 \nu ^2-452992400 \nu ^3\right) \mathbf{v}^4 \bigr]\\
& \nm
-\frac{2 G^4 M^5 \nu^2}{51975 \, r^5} \bigl[ 3 \left(-5476951+30229425 \nu -26025095 \nu ^2+4302385 \nu ^3\right) \left(\mathbf{n} \cdot \mathbf{v} \right)^6 \\
& \nm
-5 \left(-6436413+35158037 \nu -33049287 \nu ^2+6968200 \nu ^3\right)\left(\mathbf{n} \cdot \mathbf{v} \right)^4 \mathbf{v}^2 \\
& \nm
+15 \left(-1230255+6688869 \nu -6699393 \nu ^2+1862198 \nu ^3\right) \left(\mathbf{n} \cdot \mathbf{v} \right)^2 \mathbf{v}^4 \\
& \nm
-15 \left(-185007+1048531 \nu -955013 \nu ^2+378040 \nu ^3\right) \mathbf{v}^6 \bigr]\\
& \nm
+\frac{ G^3 M^4 \nu^2}{6930 \, r^4} \bigl[ 5 \left(301585+1433696 \nu + 1099688 \nu ^2-209632 \nu ^3\right) \left(\mathbf{n} \cdot \mathbf{v} \right)^8 \\
& \nm
+4 \left(-1005979+5179198 \nu -5288564 \nu ^2+1447280 \nu ^3\right)\left(\mathbf{n} \cdot \mathbf{v} \right)^6 \mathbf{v}^2 \\
& \nm
-6 \left(-613047+3522149 \nu -4709506 \nu ^2+1751152 \nu ^3\right) \left(\mathbf{n} \cdot \mathbf{v} \right)^4\mathbf{v}^4 \\
& \nm
+4 \left(-346489+2239826 \nu -3754884 \nu ^2+1858800 \nu ^3\right) \left(\mathbf{n} \cdot \mathbf{v} \right)^2\mathbf{v}^6 \bigr]\\
& \nm
+\left(240945-1388854 \nu +2416340 \nu ^2-1630560 \nu ^3\right) \mathbf{v}^8 \bigr]\\
& \nm
+\frac{6848 G^6 M^7 \nu^2}{1575 r^7}\left(\frac{2}{d-3} -5 \Log{\left( \mu_s^2 r^2\right)} \right)\left[2\left(\mathbf{n} \cdot \mathbf{v} \right)^2 -3\mathbf{v}^2 \right]\\
&
+\frac{3424 G^5 M^6 \nu^2}{525 r^6}\left(\frac{1}{d-3} -2\Log{\left( \mu_s^2 r^2\right)} \right)\left[35\left(\mathbf{n} \cdot \mathbf{v} \right)^4 -48\left(\mathbf{n} \cdot \mathbf{v} \right)^2\mathbf{v}^2+12\mathbf{v}^4 \right] \,.
\end{align}
 As anticipated, the reader will immediately notice the presence of a UV divergent term $\propto 1/(d-3)$, as well as the arbitrary renormalization scale $\mu_s$.  As we mentioned, the UV divergence and  $\mu_s$-dependence must cancel out against the hereditary contribution (from the tail-of-tail) to the (radiative) quadrupole emission. For instance, for the case of quasi-circular orbits, computing the (source) GW flux using the result in~\eqref{eq:3pn_flux}, combined with the hereditary contributions (from the tail and tail-of-tail) computed in \cite{Goldberger:2009qd},\footnote{Both the leading and NLO tail contribution to the flux were obtained using the tail's universal character (see Eqs. (12)~and~(22) in~\cite{Goldberger:2009qd}), which applies to all the relevant (mass- and current-type) multipole moments. Moreover, the (``instantaneous") contributions due to memory terms, in principle at 2.5PN order, can be shown to vanish (See Eqs. (8.2) and (8.3) in~\cite{Arun:2007sg}).}
\begin{align}\label{eq:3pn_flux_tail}
\mathcal{P}_{\text{circ}}^{(\text{hered})}
= & \nm
\frac{32 \nu ^2}{5 G} x^5 \bigg\{4\pi x^{3/2}+\left(-\frac{8191}{672} -\frac{583}{24}\nu\right)\pi x^{5/2}\\
&+\left[\frac{455153}{11025}+\frac{16 \pi^2}{3}-\frac{1712}{105}\gamma_E-\frac{856}{105}\left(\frac{1}{d-3}+4\Log 2 -\frac{7}{3} \Log x + \frac{5}{3}\Log \bar\mu^2_s \right)\right]x^3+ {\cal O}(x^{7/2})\bigg\}\,,
\end{align}
where $\bar\mu_s \equiv GM \mu_s$, we arrive at  the total energy flux, 
\begin{align}\label{eq:3pn_flux_circular}
\mathcal{P}_\text{circ}
= & \nm \mathcal{P}_{\text{circ}}^{(\text{inst})}+\mathcal{P}_{\text{circ}}^{(\text{hered})}=
\frac{32 \nu ^2}{5 G} x^5 \bigg\{ 1+ \left(-\frac{1247}{336}-\frac{35}{12}\nu\right)x +4\pi x^{3/2}\\
& \nm
+\left(-\frac{44711}{9072}+\frac{9271}{504}\nu+\frac{65}{18}\nu ^2 \right)x^2+\left(-\frac{8191}{672} -\frac{583}{24}\nu\right)\pi x^{5/2} \\
& 
+\left[\frac{6643739519}{69854400}-\frac{1712}{105} \gamma_{E}+\frac{16 \pi^2}{3}+\left(-\frac{134543}{7776}+\frac{41}{48} \pi^2\right)\nu \right.\nm \\ &-\left.\frac{94403 }{3024}\nu^2-\frac{775}{324}\nu^3-\frac{856}{105} \Log{(16 \,x)}\right ]x^3+{\cal O}(x^{7/2})\bigg\} \,.
\end{align}
 Gladly, the above expression is not only scheme-independent and UV-finite, it is also in perfect agreement with the previous 3PN result \cite{Blanchet:2001aw,Blanchet:2004ek}, see also~\cite{Blanchet:2013haa}.

\section{Conclusions}\label{sec:concl}

Using the EFT approach to gravitational dynamics, we computed the equations of motion and (source) multipole moments needed for the derivation of the (instantaneous) contribution to the GW flux at N$^3$LO. Upon including the hereditary corrections due to the tail and tail-of-tail effects obtained in \cite{Goldberger:2009qd}, the calculations presented here provide---for the first time---an independent confirmation of the value of the GW flux for circular orbits at 3PN order derived in \cite{Blanchet:2001aw,Blanchet:2004ek}. The EFT formalism readily incorporates all the relevant contributions from the ``near" and ``far" zones, yielding UV finite results without the need of the ``ambiguity parameters" that polluted the original derivations \cite{Blanchet:2001ax,Blanchet:2001aw,Blanchet:2004ek}. Our results here illustrate, similarly to the resolution of IR ambiguities in the conservative sector at 4PN order \cite{Galley:2015kus,Porto:2017dgs,Foffa:2019yfl}, the usefulness of the separation of scales and matching computation introduced in \cite{Goldberger:2004jt}, in conjunct with dim. reg. to handle divergent integrals (either due to the point-particle approximation and/or split into regions). Similarly to the upgrading to spin effects in the conservative sector, by simply including new worldline couplings \cite{Porto:2005ac,Porto:2006bt,Porto:2008jj,Porto:2008tb,Porto:2010tr,Porto:2010zg,Porto:2012as,Cho:2021mqw,Cho:2022syn,Porto:2016pyg}, the derivations presented here pave the way to the inclusion of spin corrections in the GW flux to the same N$^3$LO level of precision, up to 5PN order, that will be presented elsewhere. We also do not foresee any obstacle to continue moving forward to higher orders, towards the present state of the art at N$^4$LO~\cite{Blanchet:2023bwj,Blanchet:2023sbv}.\vskip 12pt 

{\bf Acknowledgements.} It is a pleasure to thank Fran\c cois Larrouturou for several enlightening discussions, comments, and cross checks. 
The work of L.A. was supported by the International Helmholtz-Weizmann Research School for Multimessenger Astronomy, funded through the Initiative and Networking Fund of the Helmholtz Association. The work of R.A.P and Z.Y. was funded by the ERC-CoG ``Precision Gravity: From the LHC to LISA'' provided by the European Research Council under the European Union’s H2020 research and innovation program (grant agreement No. 817791).

\appendix


\section{Post-Newtonian toolkit}\label{sec:app_2pn}

\begin{figure}[htbp]
  \centering
  \begin{subfigure}[b]{0.13\textwidth}
    \includegraphics[width=\linewidth]{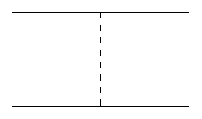}
  \end{subfigure}
  \begin{subfigure}[b]{0.13\textwidth}
    \includegraphics[width=\linewidth]{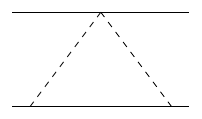}
  \end{subfigure}
  \begin{subfigure}[b]{0.13\textwidth}
    \includegraphics[width=\linewidth]{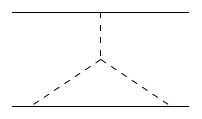}
  \end{subfigure}
  \begin{subfigure}[b]{0.13\textwidth}
    \includegraphics[width=\linewidth]{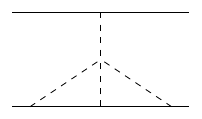}
  \end{subfigure}
  \begin{subfigure}[b]{0.13\textwidth}
    \includegraphics[width=\linewidth]{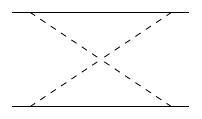}
  \end{subfigure}
    \begin{subfigure}[b]{0.13\textwidth}
    \includegraphics[width=\linewidth]{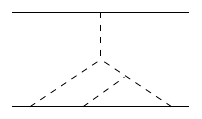}
  \end{subfigure}
  \begin{subfigure}[b]{0.13\textwidth}
    \includegraphics[width=\linewidth]{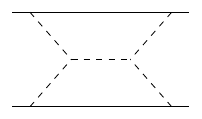}
  \end{subfigure}
  \begin{subfigure}[b]{0.13\textwidth}
    \includegraphics[width=\linewidth]{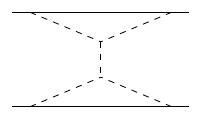}
  \end{subfigure}
    \begin{subfigure}[b]{0.13\textwidth}
    \includegraphics[width=\linewidth]{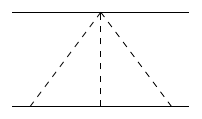}
  \end{subfigure}
    \begin{subfigure}[b]{0.13\textwidth}
    \includegraphics[width=\linewidth]{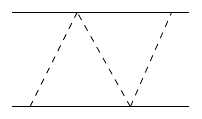}
  \end{subfigure}
    \begin{subfigure}[b]{0.13\textwidth}
    \includegraphics[width=\linewidth]{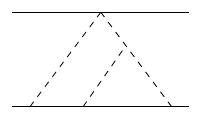}
  \end{subfigure}
    \begin{subfigure}[b]{0.13\textwidth}
    \includegraphics[width=\linewidth]{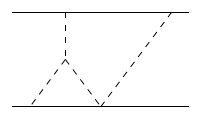}
  \end{subfigure}
  \caption{Topologies needed for the gravitational potential to N$^2$LO  (without mirror images).}
  \label{fig:bare_pot}
\end{figure}

\begin{figure}[htbp]
  \centering
  \begin{subfigure}[b]{0.2\textwidth}
    \includegraphics[width=\linewidth]{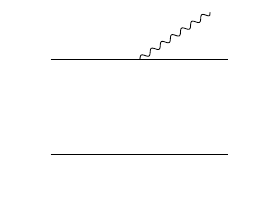}
  \end{subfigure}
  \begin{subfigure}[b]{0.2\textwidth}
    \includegraphics[width=\linewidth]{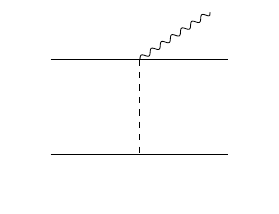}
  \end{subfigure}
  \begin{subfigure}[b]{0.2\textwidth}
    \includegraphics[width=\linewidth]{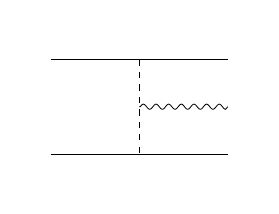}
  \end{subfigure}
  \medskip
  \begin{subfigure}[b]{0.2\textwidth}
    \includegraphics[width=\linewidth]{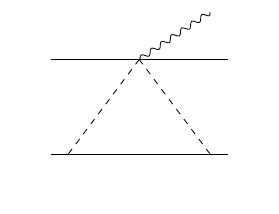}
  \end{subfigure}
  \begin{subfigure}[b]{0.2\textwidth}
    \includegraphics[width=\linewidth]{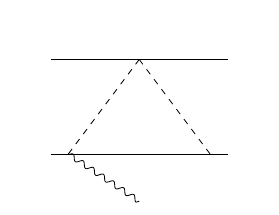}
  \end{subfigure}
  \begin{subfigure}[b]{0.2\textwidth}
    \includegraphics[width=\linewidth]{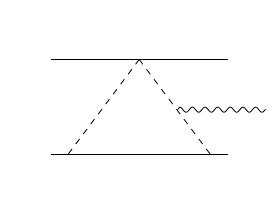}
  \end{subfigure}
  \begin{subfigure}[b]{0.2\textwidth}
    \includegraphics[width=\linewidth]{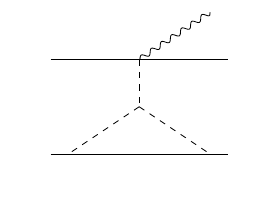}
  \end{subfigure}
  \begin{subfigure}[b]{0.2\textwidth}
    \includegraphics[width=\linewidth]{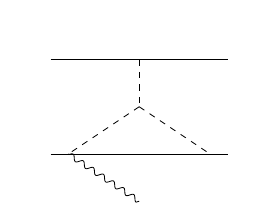}
  \end{subfigure}
    \medskip
  \begin{subfigure}[b]{0.2\textwidth}
    \includegraphics[width=\linewidth]{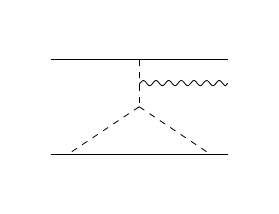}
  \end{subfigure}
  \begin{subfigure}[b]{0.2\textwidth}
    \includegraphics[width=\linewidth]{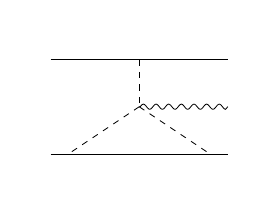}
  \end{subfigure}
  \begin{subfigure}[b]{0.2\textwidth}
    \includegraphics[width=\linewidth]{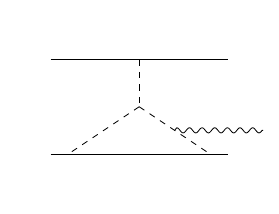}
  \end{subfigure}
  \caption{Topologies needed for the radiative sector to N$^2$LO (without mirror images).}
  \label{fig:bare_rad}
\end{figure}

This appendix lists various expressions and relations previously obtained in  the literature which are relevant for our computations here. The two set of topologies needed to 2PN order are shown in Figs.~\ref{fig:bare_pot} and~\ref{fig:bare_rad} for the conservative and radiative dynamics, respectively. The evaluation of these diagrams follows the same steps reported in the main text for the 3PN case, and only two- and one-loop master integrals are needed. While the two-loop ones are already shown in\eqref{eq:2loop_mi}, the generalized formula for the single one-loop master integral is given by
\begin{align}\label{MI_One}
M_{1,1}
& = 
\int_{\mathbf{q}_1} \frac{1}{\left[\mathbf{q}_1^2\right]^a\left[\left(\mathbf{q}_1+\mathbf{p}\right)^2\right]^b} = \frac{\left(\mathbf{p}^2\right)^{\frac{d}{2}-a-b}}{\left( 4 \pi \right)^{\frac{d}{2}}} \frac{\Gamma \left(a+b-\frac{d}{2} \right)}{\Gamma\left(a\right)\Gamma\left(b\right)} \frac{\Gamma \left(\frac{d}{2}-a \right)\Gamma\left(\frac{d}{2}-b \right)}{\Gamma\left(d-a-b\right)} \,.
\end{align}
This formula is also used for the evaluation of~\eqref{eq:three_loop} and~\eqref{eq:2loop_mi}. The final integration over the momentum transfer, $\mathbf{p}$, is always performed via the Fourier integral
\begin{equation}\label{eq:fourier_int}
\int_\mathbf{p} \frac{e^{-i \mathbf{p}\cdot\mathbf{r}}}{\left[\mathbf{p}^2\right]^a} = \frac{1}{\left( 4 \pi \right)^{\frac{d}{2}}} \frac{\Gamma\left(\frac{d}{2}-a\right)}{\Gamma\left(a\right)}\left(\frac{r^2}{4}\right)^{a-\frac{d}{2}}\,.
\end{equation}
which can also be used to derive tensorial expressions, e.g.
\begin{equation}
    \int_\mathbf{p} \frac{e^{-i \mathbf{p}\cdot\mathbf{r}}\mathbf{p}^i\mathbf{p}^j\mathbf{p}^{\ell}}{\left[\mathbf{p}^2\right]^a} = i\frac{\partial}{\partial \mathbf{r}^i}i\frac{\partial}{\partial \mathbf{r}^j}i\frac{\partial}{\partial \mathbf{r}^{\ell}}\int_\mathbf{p} \frac{e^{-i \mathbf{p}\cdot\mathbf{r}}}{\left[\mathbf{p}^2\right]^a}\,,
\end{equation}

From the equations of motion we can derive the PN corrections to the center-of-mass frame, yielding
\begin{subequations}\label{eq:com_2pn}
\begin{align}
\delta \mathbf{r}_{1\rm PN}
& = \frac{\Delta \nu}{2}\left(-\frac{GM}{r}+\mathbf{v}^2\right)\mathbf{r}\,,\\
\delta \mathbf{r}_{2\rm PN}
& = \nm 
\Delta \nu \biggl\{-\frac{G^2 M^2}{4 \, r^2}(-7+2\nu)+\frac{G M}{8 \, r}\left[(-1+6\nu)\left(\mathbf{n} \cdot \mathbf{v} \right)^2+(19+12 \nu)\mathbf{v}^2\right]\\
&
\quad +\frac{3}{8}(1-4\nu)\mathbf{v}^4\biggr\}\mathbf{r}-\frac{7\,G M}{4}\Delta \nu \left(\mathbf{n} \cdot \mathbf{v} \right) \mathbf{v}\,,\\
\delta \mathbf{v}_{1\rm PN}
& =\Delta \nu\biggl\{-\frac{G M}{2\,r^2}\left(\mathbf{n} \cdot \mathbf{v} \right)\mathbf{r}+\frac{1}{2}\left[-\frac{G M}{2\, r}+\mathbf{v}^2\right]\mathbf{v}\biggr\}\,,\\
\delta \mathbf{v}_{2 \rm PN}
& = \nm 
\Delta \nu \biggl\{-\frac{3\,G^2 M^2}{4 \, r^3}(3+2\nu)\left(\mathbf{n} \cdot \mathbf{v} \right)+\frac{G M}{8 \, r^2}\left(\mathbf{n} \cdot \mathbf{v} \right)\left[(3-6\nu)\left(\mathbf{n} \cdot \mathbf{v} \right)^2+(-9+8 \nu)\mathbf{v}^2\right]\biggr\}\mathbf{r}\\
&
\quad +\Delta \nu \biggl\{-\frac{G^2 M^2}{2 \, r^2}(-7+\nu)+\frac{G M}{8 \, r}\left[(13+6\nu)\left(\mathbf{n} \cdot \mathbf{v} \right)^2+(5+12 \nu)\mathbf{v}^2\right]+\frac{3}{8}(1-4\nu)\mathbf{v}^4\biggr\}\mathbf{v}\,,
\end{align}
\end{subequations}
which is sufficient for our purposes here. 
\section{Contributions from lower loop orders}\label{sec:app_3pn}

This appendix provides the essential resources for calculating 3PN corrections from lower order topologies, shown in Figs.~\ref{fig:bare_pot} and~\ref{fig:bare_rad}. 

\subsection{Conservative sector}\label{app:sub_cons}

\begin{figure}[htbp]
  \centering
  \begin{subfigure}[b]{0.18\textwidth}
    \includegraphics[width=\linewidth]{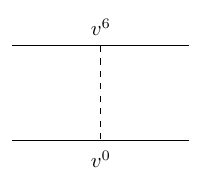}
  \end{subfigure}
  \begin{subfigure}[b]{0.18\textwidth}
    \includegraphics[width=\linewidth]{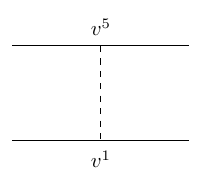}
  \end{subfigure}
  \begin{subfigure}[b]{0.18\textwidth}
    \includegraphics[width=\linewidth]{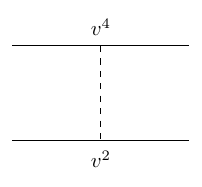}
  \end{subfigure}
  \begin{subfigure}[b]{0.18\textwidth}
    \includegraphics[width=\linewidth]{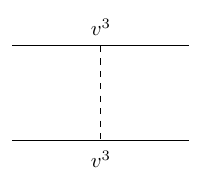}
  \end{subfigure}
  \begin{subfigure}[b]{0.18\textwidth}
    \includegraphics[width=\linewidth]{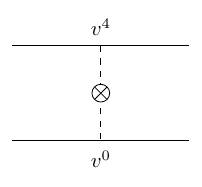}
  \end{subfigure}
  \begin{subfigure}[b]{0.18\textwidth}
    \includegraphics[width=\linewidth]{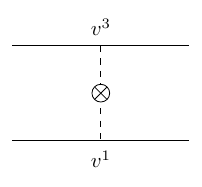}
  \end{subfigure}
    \begin{subfigure}[b]{0.18\textwidth}
    \includegraphics[width=\linewidth]{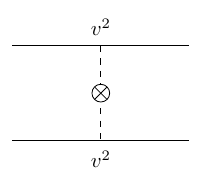}
  \end{subfigure}
  \begin{subfigure}[b]{0.18\textwidth}
    \includegraphics[width=\linewidth]{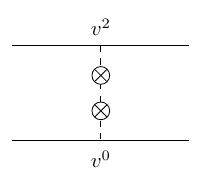}
  \end{subfigure}
  \begin{subfigure}[b]{0.18\textwidth}
    \includegraphics[width=\linewidth]{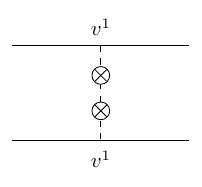}
  \end{subfigure}
    \begin{subfigure}[b]{0.18\textwidth}
    \includegraphics[width=\linewidth]{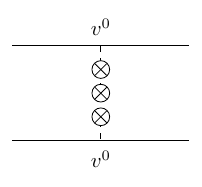}
  \end{subfigure}
      \caption{Corrections due to a single-graviton exchange at 3PN.}
  \label{fig:pot_LOex}
\end{figure}
\begin{figure}[htbp]
  \centering
  \begin{subfigure}[b]{0.16\textwidth}
    \includegraphics[width=\linewidth]{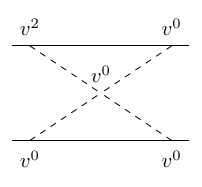}
  \end{subfigure}
  \begin{subfigure}[b]{0.16\textwidth}
    \includegraphics[width=\linewidth]{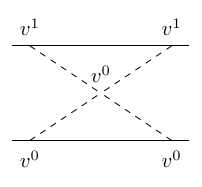}
  \end{subfigure}
  \begin{subfigure}[b]{0.16\textwidth}
    \includegraphics[width=\linewidth]{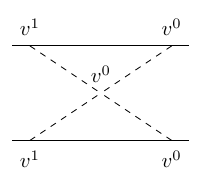}
  \end{subfigure}
  \begin{subfigure}[b]{0.16\textwidth}
    \includegraphics[width=\linewidth]{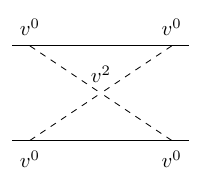}
  \end{subfigure}
  \begin{subfigure}[b]{0.16\textwidth}
    \includegraphics[width=\linewidth]{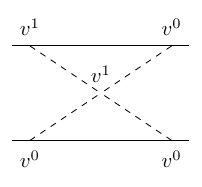}
  \end{subfigure}
  \begin{subfigure}[b]{0.16\textwidth}
    \includegraphics[width=\linewidth]{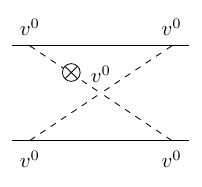}
  \end{subfigure}
  \caption{Corrections due to the four-graviton vertex at 3PN.}
  \label{fig:pot_NNLOex}
\end{figure}

The vertices detailing the couplings between worldlines and potential gravitons are given by:
\begin{subequations}\label{eq:vertex_cons1}
\begin{align}
&
S_H^{\mathbf{v}^0}=-\sum_{a=1,2} \frac{m_a}{2 m_{Pl}}\,\int \dd t_a \, H^{00}(x_a)\,,\\
&
S_H^{\mathbf{v}^1}=\sum_{a=1,2} \frac{m_a}{m_{Pl}}\,\int \dd t_a \, \mathbf{v}_a^i \, H^{0i}(x_a)\,,\\
&
S_H^{\mathbf{v}^2}=-\sum_{a=1,2} \frac{m_a}{2 m_{Pl}}\,\int \dd t_a \left(\frac{\mathbf{v}_a^2}{2}\, H^{00}(x_a)+\mathbf{v}_a^i\mathbf{v}_a^j\,H^{ij}(x_a)\right)\,,\\
&
S_H^{\mathbf{v}^3}=\sum_{a=1,2} \frac{m_a}{2 m_{Pl}}\,\int \dd t_a \, \mathbf{v}_a^2\mathbf{v}_a^i \, H^{0i}(x_a)\,,\\
&
S_H^{\mathbf{v}^4}=-\sum_{a=1,2} \frac{m_a}{4 m_{Pl}}\,\int \dd t_a \left(\frac{3 \mathbf{v}_a^4}{4}\,H^{00}(x_a)+\mathbf{v}_a^2\mathbf{v}_a^i\mathbf{v}_a^j\,H^{ij}(x_a)\right)\,,\\
&
S_H^{\mathbf{v}^5}=\sum_{a=1,2} \frac{3m_a}{8 m_{Pl}}\,\int \dd t_a \, \mathbf{v}_a^4\mathbf{v}_a^i \, H^{0i}(x_a)\,,\\
&
S_H^{\mathbf{v}^6}=-\sum_{a=1,2} \frac{m_a}{16 m_{Pl}}\,\int \dd t_a \left(\frac{5 \mathbf{v}_a^6}{2}\,H^{00}(x_a)+3\mathbf{v}_a^4\mathbf{v}_a^i\mathbf{v}_a^j\,H^{ij}(x_a)\right)\,.
\end{align}
\end{subequations}
\begin{subequations}\label{eq:vertex_cons2}
\begin{align}
&
S_{H^2}^{\mathbf{v}^0}=\sum_{a=1,2} \frac{m_a}{8 m^2_{Pl}}\,\int \dd t_a \, H^{00}(x_a)H^{00}(x_a)\,,\\
&
S_{H^2}^{\mathbf{v}^1}=-\sum_{a=1,2} \frac{m_a}{2 m^2_{Pl}}\,\int \dd t_a \, \mathbf{v}_a^i \, H^{0i}(x_a)H^{00}(x_a)\,,\\
& \nm
S_{H^2}^{\mathbf{v}^2}=\sum_{a=1,2} \frac{m_a}{2 m^2_{Pl}}\,\int \dd t_a \biggl( \frac{3\mathbf{v}_a^2}{8}\,H^{00}(x_a)H^{00}(x_a) +\mathbf{v}^i\mathbf{v}^j\,H^{0i}(x_a)H^{0j}(x_a)\\
&
\qquad \quad+\frac{1}{2}\,\mathbf{v}_a^i\mathbf{v}_a^j\,H^{ij}(x_a)H^{00}(x_a)\biggr)\,,\\
&
S_{H^2}^{\mathbf{v}^3}=-\sum_{a=1,2} \frac{m_a}{2 m^2_{Pl}}\,\int \dd t_a \left( \frac{3}{2}\,\mathbf{v}_a^2\mathbf{v}_a^i \, H^{0i}(x_a)H^{00}(x_a)+\mathbf{v}_a^i\mathbf{v}_a^j\mathbf{v}_a^k\,H^{0i}(x_a)H^{ij}(x_a)\right)\,,\\
& \nm
S_{H^2}^{\mathbf{v}^4}=\sum_{a=1,2} \frac{m_a}{4 m^2_{Pl}}\,\int \dd t_a \biggl( \frac{15\mathbf{v}_a^4}{16}\,H^{00}(x_a)H^{00}(x_a) +3\mathbf{v}_a^2\mathbf{v}^i\mathbf{v}^j\,H^{0i}(x_a)H^{0j}(x_a)\,,\\
&
\qquad \quad+\frac{3}{2}\,\mathbf{v}_a^2\mathbf{v}_a^i\mathbf{v}_a^j\,H^{ij}(x_a)H^{00}+\frac{1}{2}\,\mathbf{v}_a^i\mathbf{v}_a^j\mathbf{v}_a^k\mathbf{v}_a^l\,H^{ij}(x_a)H^{kl}(x_a)\biggr)\,.
\end{align}
\end{subequations}
\begin{subequations}\label{eq:vertex_cons3}
\begin{align}
&
S_{H^3}^{\mathbf{v}^0}=-\sum_{a=1,2} \frac{m_a}{16 m^3_{Pl}}\,\int \dd t_a \, H^{00}(x_a)H^{00}(x_a)H^{00}(x_a)\,,\\
&
S_{H^3}^{\mathbf{v}^1}=\sum_{a=1,2} \frac{3 m_a}{8 m^3_{Pl}}\,\int \dd t_a \,\mathbf{v}_a^i H^{0i}(x_a)H^{00}(x_a)H^{00}(x_a)\,,\\
& \nm
S_{H^3}^{\mathbf{v}^2}=-\sum_{a=1,2} \frac{m_a}{4 m^3_{Pl}}\,\int \dd t_a \biggl( \frac{5\mathbf{v}_a^2}{8}\,H^{00}(x_a)H^{00}(x_a)H^{00}(x_a) +3\mathbf{v}^i\mathbf{v}^j\,H^{0i}(x_a)H^{0j}(x_a)H^{00}(x_a)\\
&
\qquad \quad+\frac{3}{4}\,\mathbf{v}_a^i\mathbf{v}_a^j\,H^{ij}(x_a)H^{00}(x_a)H^{00}(x_a)\biggr)\,.
\end{align}
\end{subequations}
\begin{equation}
    S_{H^4}^{\mathbf{v}^0}=\sum_{a=1,2} \frac{5 m_a}{128 m^4_{Pl}}\,\int \dd t_a \, H^{00}(x_a)H^{00}(x_a)H^{00}(x_a)H^{00}(x_a)\,,
\end{equation}
The Feynman rules for gravitational self-interactions of potential modes are extracted using standard packages, and with the help of the functional derivative (in mixed Fourier space)
\begin{equation}\label{eq:funct_der}
\frac{\delta H_{\alpha \beta}(t_1,\mathbf{q}_1)}{\delta H^{\mu \nu}(t_2,\mathbf{q}_2)}=\delta^{(3)}\left(\mathbf{q}_1+\mathbf{q}_2\right)\delta\left(t_1-t_2\right) I_{\mu \nu \alpha \beta}\,,
\end{equation}
where $I_{\mu \nu \alpha \beta} =\frac{1}{2}\left(\eta_{\mu \alpha}\eta_{\nu \beta}+\eta_{\mu \beta}\eta_{\nu \alpha}\right)$. Each vertex is also PN expanded, depending on the number of time derivatives. 
\begin{figure}[t!]
  \centering
  \begin{subfigure}[b]{0.16\textwidth}
    \includegraphics[width=\linewidth]{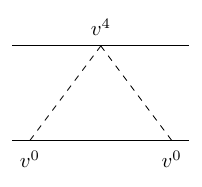}
  \end{subfigure}
  \begin{subfigure}[b]{0.16\textwidth}
    \includegraphics[width=\linewidth]{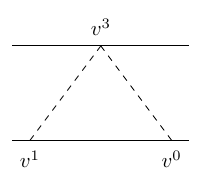}
  \end{subfigure}
  \begin{subfigure}[b]{0.16\textwidth}
    \includegraphics[width=\linewidth]{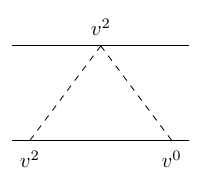}
  \end{subfigure}
  \begin{subfigure}[b]{0.16\textwidth}
    \includegraphics[width=\linewidth]{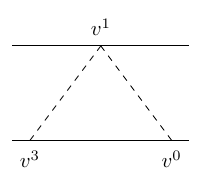}
  \end{subfigure}
  \begin{subfigure}[b]{0.16\textwidth}
    \includegraphics[width=\linewidth]{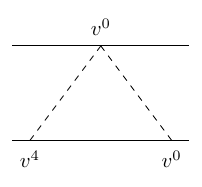}
  \end{subfigure}
  \begin{subfigure}[b]{0.16\textwidth}
    \includegraphics[width=\linewidth]{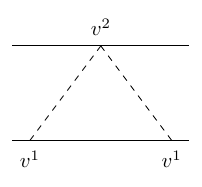}
  \end{subfigure}
    \begin{subfigure}[b]{0.16\textwidth}
    \includegraphics[width=\linewidth]{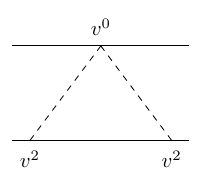}
  \end{subfigure}
  \begin{subfigure}[b]{0.16\textwidth}
    \includegraphics[width=\linewidth]{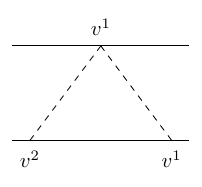}
  \end{subfigure}
  \begin{subfigure}[b]{0.16\textwidth}
    \includegraphics[width=\linewidth]{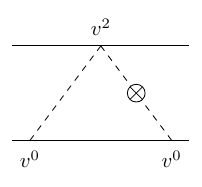}
  \end{subfigure}
  \begin{subfigure}[b]{0.16\textwidth}
    \includegraphics[width=\linewidth]{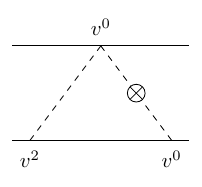}
  \end{subfigure}
    \begin{subfigure}[b]{0.16\textwidth}
    \includegraphics[width=\linewidth]{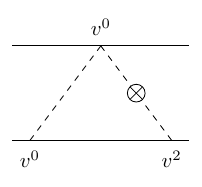}
  \end{subfigure}
  \begin{subfigure}[b]{0.16\textwidth}
    \includegraphics[width=\linewidth]{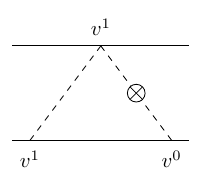}
  \end{subfigure}
    \begin{subfigure}[b]{0.16\textwidth}
    \includegraphics[width=\linewidth]{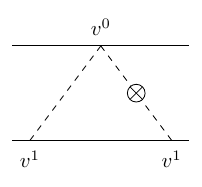}
  \end{subfigure}
  \begin{subfigure}[b]{0.16\textwidth}
    \includegraphics[width=\linewidth]{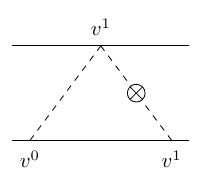}
  \end{subfigure}
  \begin{subfigure}[b]{0.16\textwidth}
    \includegraphics[width=\linewidth]{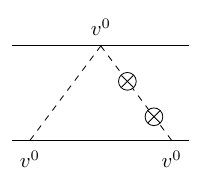}
  \end{subfigure}
  \begin{subfigure}[b]{0.16\textwidth}
    \includegraphics[width=\linewidth]{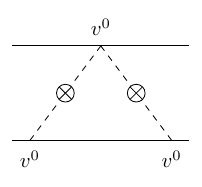}
  \end{subfigure}
  \caption{Corrections due to ``seagull'' diagrams at 3PN.}
  \label{fig:pot_NLOex}
\end{figure}
The PN corrections due to the non-instantaneous nature of the binding potentials is obtained by expanding the propagator, and is represented with a cross for each insertion of a factor of $q_0^2/\mathbf{q}^2$, i.e. 
\begin{align}
\frac{1}{q_0^2-\mathbf{q}^2} 
& \nm
\,\simeq -\frac{1}{\mathbf{q}^2}\left[1+\frac{q_0^2}{\mathbf{q}^2}+\frac{q_0^4}{\mathbf{q}^4}+\cdots\right]\\
& 
\includegraphics[width=0.7\textwidth]{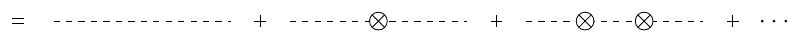}
\end{align}
The contributions from a mixture of non-linear gravitational effects, at various PN orders, together with non-instantaneous corrections to the propagators, are depicted in Figs.~\ref{fig:pot_LOex}, \ref{fig:pot_NLOex}, and \ref{fig:pot_NNLOex} 

\subsection{Radiative sector}\label{app:sub_rad}

\begin{figure}[htbp]
  \centering
  \begin{subfigure}[b]{0.18\textwidth}
    \includegraphics[width=\linewidth]{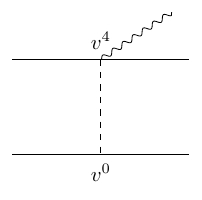}
  \end{subfigure}
  \begin{subfigure}[b]{0.18\textwidth}
    \includegraphics[width=\linewidth]{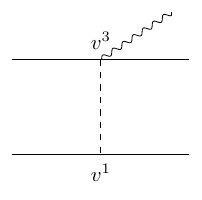}
  \end{subfigure}
  \begin{subfigure}[b]{0.18\textwidth}
    \includegraphics[width=\linewidth]{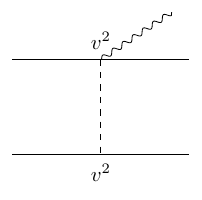}
  \end{subfigure}
  \begin{subfigure}[b]{0.18\textwidth}
    \includegraphics[width=\linewidth]{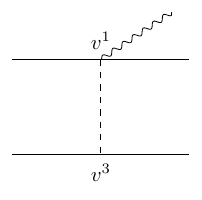}
  \end{subfigure}
  \begin{subfigure}[b]{0.18\textwidth}
    \includegraphics[width=\linewidth]{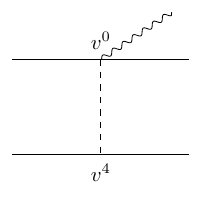}
  \end{subfigure}
  \medskip
  \begin{subfigure}[b]{0.18\textwidth}
    \includegraphics[width=\linewidth]{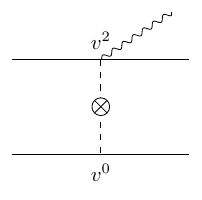}
  \end{subfigure}
  \begin{subfigure}[b]{0.18\textwidth}
    \includegraphics[width=\linewidth]{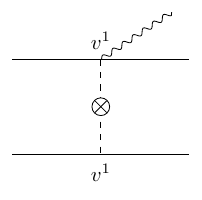}
  \end{subfigure}
  \begin{subfigure}[b]{0.18\textwidth}
    \includegraphics[width=\linewidth]{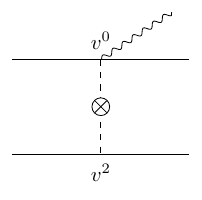}
  \end{subfigure}
  \begin{subfigure}[b]{0.18\textwidth}
    \includegraphics[width=\linewidth]{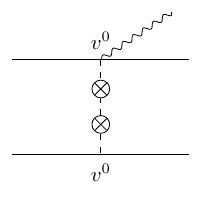}
  \end{subfigure}
  \caption{Corrections due to a single-graviton exchange radiative effects at 3PN}
    \label{fig:rad_NLOex}
\end{figure}
\begin{figure}[htbp]
  \centering
  \begin{subfigure}[b]{0.18\textwidth}
    \includegraphics[width=\linewidth]{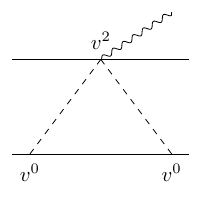}
  \end{subfigure}
  \begin{subfigure}[b]{0.18\textwidth}
    \includegraphics[width=\linewidth]{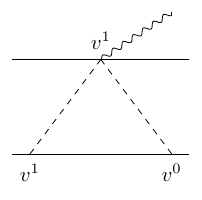}
  \end{subfigure}
  \begin{subfigure}[b]{0.18\textwidth}
    \includegraphics[width=\linewidth]{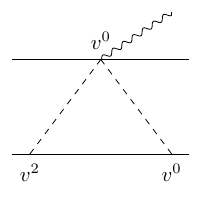}
  \end{subfigure}
  \begin{subfigure}[b]{0.18\textwidth}
    \includegraphics[width=\linewidth]{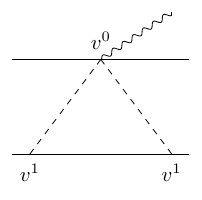}
  \end{subfigure}
  \begin{subfigure}[b]{0.18\textwidth}
    \includegraphics[width=\linewidth]{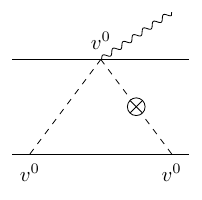}
  \end{subfigure}
  \caption{Corrections due to seagull-type radiative effects at 3PN.}
    \label{fig:rad_NNLOex}
\end{figure}

Following the same logic as in the conservative sector, the worldline couplings to both potential and radiation fields are listed below:
\begin{subequations}\label{eq:vertex_rad1}
\begin{align}
&
S_{H \bar{h}}^{\mathbf{v}^0}=\sum_{a=1,2} \frac{m_a}{4 m^2_{Pl}}\,\int \dd t_a \, H^{00}(x_a)\bar{h}^{00}(x_a)\,,\\
&
S_{H \bar{h}}^{\mathbf{v}^1}=-\sum_{a=1,2} \frac{m_a}{2 m^2_{Pl}}\,\int \dd t_a \, \mathbf{v}_a^i \left[ H^{0i}(x_a)\bar{h}^{00}(x_a)+H^{00}(x_a) \bar{h}^{0i}(x_a)\right]\,,\\
& \nm
S_{H \bar{h}}^{\mathbf{v}^2}=\sum_{a=1,2} \frac{m_a}{m^2_{Pl}}\,\int \dd t_a \biggl\{ \frac{3\mathbf{v}_a^2}{8}\,H^{00}(x_a)\bar{h}^{00}(x_a) +\mathbf{v}^i\mathbf{v}^j\, H^{0i}(x_a)\bar{h}^{0j}(x_a)\\
&
\qquad \quad+\frac{1}{4}\,\mathbf{v}_a^i\mathbf{v}_a^j\left[H^{ij}(x_a)\bar{h}^{00}(x_a)+H^{00}(x_a)\bar{h}^{ij}(x_a)\right]\biggr\}\,,\\
& \nm
S_{H \bar{h}}^{\mathbf{v}^3}=-\sum_{a=1,2} \frac{m_a}{2 m^2_{Pl}}\,\int \dd t_a \biggl\{ \frac{3}{2}\,\mathbf{v}_a^2\mathbf{v}_a^i \left[ H^{0i}(x_a)\bar{h}^{00}(x_a)+H^{00}(x_a)\bar{h}^{0i}(x_a)\right]\\
&
\qquad \quad+\mathbf{v}_a^i\mathbf{v}_a^j\mathbf{v}_a^k\left[H^{0i}(x_a)\bar{h}^{ij}(x_a)+H^{ij}(x_a)\bar{h}^{0i}(x_a)\right]\biggr\}\,,\\
& \nm
S_{H \bar{h}}^{\mathbf{v}^4}=\sum_{a=1,2} \frac{m_a}{2 m^2_{Pl}}\,\int \dd t_a \biggl\{ \frac{15\mathbf{v}_a^4}{16}\,H^{00}(x_a)\bar{h}^{00}(x_a) +3\mathbf{v}_a^2\mathbf{v}^i\mathbf{v}^j\,H^{0i}(x_a)\bar{h}^{0j}(x_a)\,,\\
&
\qquad \quad+\frac{3}{4}\,\mathbf{v}_a^2\mathbf{v}_a^i\mathbf{v}_a^j\left[H^{ij}(x_a)\bar{h}^{00}+H^{00}(x_a)\bar{h}^{ij}\right]+\frac{1}{2}\,\mathbf{v}_a^i\mathbf{v}_a^j\mathbf{v}_a^k\mathbf{v}_a^l\,H^{ij}(x_a)\bar{h}^{kl}(x_a)\biggr\}\,.
\end{align}
\end{subequations}
\begin{subequations}\label{eq:vertex_rad2}
\begin{align}
&
S_{H^2 \bar{h}}^{\mathbf{v}^0}=-\sum_{a=1,2} \frac{3 m_a}{16 m^3_{Pl}}\,\int \dd t_a \, H^{00}(x_a)H^{00}(x_a)\bar{h}^{00}(x_a)\,,\\
&
S_{H^2 \bar{h}}^{\mathbf{v}^1}=\sum_{a=1,2} \frac{3 m_a}{4 m^3_{Pl}}\,\int \dd t_a \,\mathbf{v}_a^i \left[H^{0i}(x_a)H^{00}(x_a)\bar{h}^{00}(x_a)+\frac{1}{2}\,H^{00}(x_a)H^{00}(x_a)\bar{h}^{0i}(x_a)\right]\,,\\
& \nm
S_{H^2 \bar{h}}^{\mathbf{v}^2}=-\sum_{a=1,2} \frac{m_a}{2 m^3_{Pl}}\,\int \dd t_a \biggl\{ \frac{15\mathbf{v}_a^2}{16}\,H^{00}(x_a)H^{00}(x_a)\bar{h}^{00}(x_a) \\
& \nm
\qquad \quad+3\mathbf{v}^i\mathbf{v}^j\left[H^{00}(x_a)H^{0i}(x_a)\bar{h}^{0j}(x_a)+\frac{1}{2}\,H^{0i}(x_a)H^{0j}(x_a)\bar{h}^{00}(x_a)\right]\\
&
\qquad \quad+\frac{3}{4}\,\mathbf{v}_a^i\mathbf{v}_a^j\left[H^{ij}(x_a)H^{00}(x_a)\bar{h}^{00}(x_a)+\frac{1}{2}\,H^{00}(x_a)H^{00}(x_a)\bar{h}^{ij}(x_a)\right]\biggr\}\,,
\end{align}
\end{subequations}
\begin{equation}\label{eq:vertex_rad3}
S_{H^3 \bar{h}}^{\mathbf{v}^0}=\sum_{a=1,2} \frac{5 m_a}{32 m^4_{Pl}}\,\int \dd t_a \, H^{00}(x_a)H^{00}(x_a)H^{00}(x_a)\bar{h}^{00}(x_a)\,.
\end{equation}
The Feynamn rules are again derived with the help of~\eqref{eq:funct_der}. The Feynman diagrams needed at NLO and N$^2$LO are shown in Figs.~\ref{fig:rad_NLOex} and~\ref{fig:rad_NNLOex}, respectively, with the corresponding PN corrections needed to 3PN order.

\subsection{Remaining multipoles for the 3PN energy flux}\label{appb3}

In the main part of this paper we focused on the computation of the mass quadrupole at N$^3$LO. However, other multipole moments are also needed to complete the knowledge of the 3PN energy flux in~\eqref{eq:3pn_flux_symbolic}. We briefly describe below the (re)derivation of the necessary multipole moments, and the comparison to analogous results available in the existing literature. See ancillary file for computer-readable expressions.\vskip 4pt
Other than the mass quadrupole, we also require the mass octupole, as well as the current quadrupole, needed to 2PN order,
\begin{subequations}
\begin{align}
&I^{ijk}_{2\rm PN}
=  \nm
\left[\int\!\! \dd^3 \mathbf{x}\, T^{00}_{2\rm PN}\mathbf{x}^i\mathbf{x}^j\mathbf{x}^k\right]_{\text{TF}}+\left[\int\!\! \dd^3 \mathbf{x}\, T^{a a}_{1\rm PN}\mathbf{x}^i\mathbf{x}^j\mathbf{x}^k\right]_{\text{TF}} - \left[\int\!\! \dd^3 \mathbf{x}\, \partial_t\tilde{T}^{0}_{1\rm PN}\mathbf{x}^i\mathbf{x}^j\mathbf{x}^k\right]_{\text{TF}}\\
& \nm
\qquad \quad \!+\frac{1}{10}\left[\int\!\! \dd^3 \mathbf{x}\, \partial_t^2\tilde{T}_{0\rm PN}\mathbf{x}^i\mathbf{x}^j\mathbf{x}^k\right]_{\text{TF}}+\frac{1}{6}\left[\int\!\! \dd^3 \mathbf{x}\, \partial_t^2T^{00}_{1\rm PN} r^2 \mathbf{x}^i\mathbf{x}^j\mathbf{x}^k\right]_{\text{TF}}+\frac{1}{15}\left[\int\!\! \dd^3 \mathbf{x}\, \partial_t^2 T^{kk}_{0\rm PN} r^2 \mathbf{x}^i\mathbf{x}^j\mathbf{x}^k\right]_{\text{TF}}\\
& \nm
\qquad \quad \!- \frac{7}{90}\left[\int\!\! \dd^3 \mathbf{x}\, \partial_t^3 \tilde{T}^{0}_{0\rm PN} r^2 \mathbf{x}^i\mathbf{x}^j\mathbf{x}^k\right]_{\text{TF}}+ \frac{29}{3960}\left[\int\!\! \dd^3 \mathbf{x}\, \partial_t^4 T^{00}_{0\rm PN} r^4 \mathbf{x}^i\mathbf{x}^j\mathbf{x}^k\right]_{\text{TF}}\\
& 
\qquad \quad \!+ \text{lower order corrections}\\ 
&
J^{ij}_{2\rm PN}
=  \nm
\left[\int\!\! \dd^3 \mathbf{x}\, \varepsilon^{iba}T^{0a}_{2\rm PN}\mathbf{x}^b\mathbf{x}^j\right]_{\text{STF}}-\frac{1}{4}\left[\int\!\! \dd^3 \mathbf{x}\, \varepsilon^{iba}\partial_t\tilde{T}^{a}_{1\rm PN}\mathbf{x}^b\mathbf{x}^j\right]_{\text{STF}} +\frac{3}{28}\left[\int\!\! \dd^3 \mathbf{x}\, \varepsilon^{iba}\partial^2_t T^{0a}_{1\rm PN} r^2 \mathbf{x}^b\mathbf{x}^j\right]_{\text{STF}}\\
& \nm
\qquad \quad \!-\frac{1}{56}\left[\int\!\! \dd^3 \mathbf{x}\, \varepsilon^{iba}\partial^3_t\tilde{T}^{a}_{0\rm PN} r^2\mathbf{x}^b\mathbf{x}^j\right]_{\text{STF}} +\frac{1}{252}\left[\int\!\! \dd^3 \mathbf{x}\, \varepsilon^{iba}\partial^4_t T^{0a}_{1\rm PN} r^4 \mathbf{x}^b\mathbf{x}^j\right]_{\text{STF}}\\
&
\qquad \quad \!+ \text{lower order corrections}\,,
\end{align}
\end{subequations}
in terms of moments of the stress-energy tensor. Similarly to the 2PN corrections to the mass quadrupole, calculated in \cite{Leibovich:2019cxo}, all the relevant coefficients are extracted from the topologies shown in Fig.~\ref{fig:bare_rad}.  Upon performing the matching computation, our expression for the mass octupole and current quadrupole, yield
\begin{align}\label{eq:2pn_mo}
I^{ijk}_{\rm 2PN}\left(\mathbf{r},\mathbf{v}\right)
= & \nm
\biggl\{\biggl\{-\frac{G^2 M^3 \Delta \nu}{132\, r^2}\left(604-1591 \nu +470 \nu ^2\right)\\
 & \nm
+\frac{G M^2 \Delta \nu}{1320 \, r} \bigl[\left(247-1593 \nu +4041 \nu ^2\right)\left(\mathbf{n} \cdot \mathbf{v} \right)^2 +\left(-3853+14257 \nu +17371 \nu ^2\right)\mathbf{v}^2\bigr]\\
& \nm
-\frac{M \Delta \nu}{1320} \left(771-7319 \nu +16503 \nu ^2\right) \mathbf{v}^4 \biggr\}\mathbf{r}^i\mathbf{r}^j\mathbf{r}^k\\
& \nm
+\biggl\{-\frac{G M^2 \Delta \nu}{660} \left(-2461+8689 \nu +4167 \nu ^2\right)\left(\mathbf{n} \cdot \mathbf{v} \right) 
+\frac{M \Delta \nu \,r}{22}\left(13-107 \nu+204\nu^2\right)\left(\mathbf{n} \cdot \mathbf{v} \right)\mathbf{v}^2\biggr\}\mathbf{r}^i\mathbf{r}^j\mathbf{v}^k\\
& \nm
+\biggl\{\frac{G M^2 \Delta \nu \, r}{330} \left(-1949-124 \nu +2898 \nu ^2\right) 
+\frac{M \Delta \nu \,r^2}{110}\left[10\left(-1+ \nu\right) \left(\mathbf{n} \cdot \mathbf{v} \right)^2+\left(61-336 \nu\right)\mathbf{v}^2\right]\biggr\}\mathbf{r}^i\mathbf{v}^j\mathbf{v}^k \\
&
-\frac{13\, M \Delta \nu \, r^3}{55} \left(1-4 \nu +3 \nu ^2\right)\left(\mathbf{n} \cdot \mathbf{v} \right) \mathbf{v}^i\mathbf{v}^j\mathbf{v}^k\biggr\}_{\text{STF}}\,,
\end{align}
and
\begin{align}\label{eq:2pn_cq}
J^{ij}_{\rm 2PN}\left(\mathbf{r},\mathbf{v}\right)
= & \nm
\biggl\{\biggl\{-\frac{G^2 M^3 \Delta \nu}{504\, r^2}\left(1469-3086 \nu +879 \nu ^2\right)\\
 & \nm
+\frac{G M^2 \Delta \nu}{504 \, r} \bigl[\left(5+241 \nu +1005 \nu ^2\right)\left(\mathbf{n} \cdot \mathbf{v} \right)^2 +\left(-671+2594 \nu +2541 \nu ^2\right)\mathbf{v}^2\bigr]\\
& \nm
-\frac{M \Delta \nu}{336} \left(58-616 \nu +1515 \nu ^2\right) \mathbf{v}^4 \biggr\}\left(\varepsilon^{ipq}\mathbf{r}^j+\varepsilon^{jpq}\mathbf{r}^i\right)\mathbf{r}^p\mathbf{v}^q\\
& \nm
+\biggl\{\frac{G M^2 \Delta \nu}{504} \left(-412-674 \nu +519 \nu ^2\right)\left(\mathbf{n} \cdot \mathbf{v} \right) \\
&
-\frac{25 \,M \Delta \nu \,r}{336}\left(1-7 \nu+12\nu^2\right)\left(\mathbf{n} \cdot \mathbf{v} \right)\mathbf{v}^2\biggr\}\left(\varepsilon^{ipq}\mathbf{v}^j+\varepsilon^{jpq}\mathbf{v}^i\right)\mathbf{r}^p\mathbf{v}^q\biggr\}_{\text{STF}}\,,
\end{align}
in the center-of-mass frame.\vskip 4pt

As a direct cross check of these results, we compute their values for the case of circular orbits, and implement the following coordinate shift \cite{Leibovich:2019cxo}
\begin{equation}
    \mathbf{r}_{\rm EFT} \to {\mathbf{r}}-\frac{2 G_N^2 M^2}{r^2} {\mathbf{r}}\,,
\end{equation}
such that
\begin{subequations}
\begin{align}
    I^{ijk}&=-M \Delta \nu \left\{1-\gamma \nu +\gamma^2\left(-\frac{139}{330}-\frac{11923}{660}\nu-\frac{29}{110}\nu^2\right)\right\}\left[\mathbf{r}^i\mathbf{r}^j\mathbf{r}^k\right]_{\rm STF} \nm \\
    & \quad \, -M \Delta \nu r^2 \left\{1-2\nu+\gamma\left(\frac{1066}{165}-\frac{1433}{330}\nu+\frac{21}{55}\nu^2\right)\right\}\left[\mathbf{r}^i\mathbf{r}^j\mathbf{v}^k\right]_{\rm STF} \\
    J^{ij}&=-M \Delta \nu \left\{1+\gamma\left(\frac{67}{28}-\frac{2}{7}\nu\right) +\gamma^2\left(\frac{13}{9}-\frac{4651}{252}\nu-\frac{\nu^2}{168}\right)\right\}\left[\varepsilon^{ipq}\mathbf{r}^j\mathbf{r}^p\mathbf{v}^q\right]_{\rm STF}\,,
\end{align}
\end{subequations}
where we have introduced the parameter $\gamma\equiv GM/r$. These results are in perfect agreement with those displayed \cite{Blanchet:2013haa}.\vskip 4pt

The derivation of the remaining (higher order) multipoles, needed at first Post-Newtonian and leading order, yields (for circular orbits)
\begin{subequations}
\begin{align}
    I^{ijkl}&=M  \nu \left\{1-3\nu+\gamma\left( \frac{3}{110}-\frac{25}{22}\nu+\frac{69}{22}\nu^2\right) \right\}\left[\mathbf{r}^i\mathbf{r}^j\mathbf{r}^k\mathbf{r}^l\right]_{\rm STF} \nm \\
    & \quad \, +M \nu r^2 \left\{\frac{78}{55}\left(1-5\nu+5\nu^2\right)\right\}\left[\mathbf{r}^i\mathbf{r}^j\mathbf{v}^k\mathbf{v}^l\right]_{\rm STF} \\
    I^{ijklm}&=-M \Delta \nu (1-2\nu)\left[\mathbf{r}^i\mathbf{r}^j\mathbf{r}^k\mathbf{r}^l\mathbf{r}^m\right]_{\rm STF} \\
    J^{ijk}&=M  \nu \left\{1-3\nu+\gamma\left(\frac{181}{90}-\frac{109}{18}\nu +\frac{13}{18}\nu^2\right) \right\}\left[\varepsilon^{ipq}\mathbf{r}^j\mathbf{r}^k\mathbf{r}^p\mathbf{v}^q\right]_{\rm STF} \nm \\
    &
    \quad \, +M  \nu r^2 \left\{\frac{7}{45}\left(1-5\nu+5\nu^2\right) \right\}\left[\varepsilon^{ipq}\mathbf{r}^p\mathbf{v}^j\mathbf{v}^k\mathbf{v}^q\right]_{\rm STF} \\
    J^{ijkm}&=-M \Delta \nu \left(1-2\nu\right)\left[\varepsilon^{ipq}\mathbf{r}^j\mathbf{r}^k\mathbf{r}^l\mathbf{r}^p\mathbf{v}^q\right]_{\rm STF}\,,
\end{align}
\end{subequations}
also in agreement with the results in \cite{Blanchet:2013haa}. 

\subsection{Consistency checks}\label{app:consistency}
We have also performed a few additional consistency checks. 
Firstly, we rederive the total mechanical energy of the system, which can be extracted from the purely temporal component of the pseudo stress-energy tensor in~\eqref{eq:matching_stress}, at ${\cal O}(\mathbf{k}^0)$. We find 
\begin{align}\label{eq:consistency_t003pn}
E_{2\rm PN}\left(\mathbf{r},\mathbf{v}\right)=\int\!\! \dd^3 \mathbf{x}\, T^{00}_{3\rm PN}
= 
& \nm
-\frac{3 \, G^3 M^4 \nu}{4 \, r^3} (-2+5 \nu )+\frac{G^2 M^3 \nu}{8 \, r^2} \bigl[4 +\left(-28+69 \nu +12 \nu ^2\right)\left(\mathbf{n} \cdot \mathbf{v} \right)^2 \\
& \nm
+\left(30-55 \nu +4 \nu ^2\right)\mathbf{v}^2\bigr]\\
& \nm
+\frac{G M^2 \nu}{8 \, r} \bigl\{-8+ 3 \nu  (-1+3 \nu ) \left(\mathbf{n} \cdot \mathbf{v} \right)^4 +2 \nu\left(\mathbf{n} \cdot \mathbf{v} \right)^2\left[2+ (1-15 \nu )\mathbf{v}^2\right] \\
& \nm
+\mathbf{v}^2\left[4 (3+\nu )+ \left(21-23 \nu -27 \nu ^2\right) \mathbf{v}^2\right] \bigr\}\\
&
+\frac{1}{16} M \nu  \left[8+(6-18 \nu )\mathbf{v}^2 +5\left(1-7 \nu +13 \nu ^2\right)\mathbf{v}^4\right]\,,
\end{align}
which is consistent with the 2PN corrections derived via the Legendre transformation of the Lagrangian described in~\eqref{eq:ham_formula}.\vskip 4pt The second self-consistent check is provided by the moment relation
\begin{equation}\label{eq:cons_test}
    \int\!\! \dd^3 \mathbf{x}\, T^{aa}_{2 \rm PN}=\frac{1}{2} \frac{d^2}{dt^2}\int\!\! \dd^3 \mathbf{x}\, T^{00}_{2 \rm PN}r^2\,,
\end{equation}
Indeed, the left and right-hand sides of the equation can be independently calculated, resulting in
\begin{align}\label{eq:consistency_tll}
\int\!\! \dd^3 \mathbf{x}\, T^{aa}_{2\rm PN}
= & \nm
-\frac{G^3 M^4 \nu}{4 \, r^3} (-14+15 \nu ) +\frac{G^2 M^3 \nu}{8 \, r^2} \left[20+9 (5+16 \nu )\left(\mathbf{n} \cdot \mathbf{v} \right)^2 -(17+114 \nu )\mathbf{v}^2\right]\\
& \nm
+\frac{G M^2 \nu}{8 \, r}\bigl\{-8+3 \left(-3+5 \nu +3 \nu ^2\right)\left(\mathbf{n} \cdot \mathbf{v} \right)^4+\left(10+15 \nu -27 \nu ^2\right)\mathbf{v}^4+4\nu\mathbf{v}^2\\
& \nm
+\left(\mathbf{n} \cdot \mathbf{v} \right)^2\left[4 (3+\nu )+\left(20-54 \nu -30 \nu ^2\right)\mathbf{v}^2\right]\bigr\}\\
&
+\frac{M \nu \, \mathbf{v}^2}{8} \left[8+(4-12 \nu )\mathbf{v}^2 +3 \left(1-7 \nu +13 \nu ^2\right)\mathbf{v}^4\right]\,,
\end{align}
and
\begin{align}\label{eq:consistency_t002pn}
\frac{1}{2} \frac{d^2}{dt^2}\int\!\! \dd^3 \mathbf{x}\, T^{00}_{2\rm PN}r^2
= & \nm
\frac{1}{2} \frac{d^2}{dt^2}\biggl\{G^2 M^3 \nu \left\{\frac{1}{4}\left(67-28 \nu +8 \nu ^2\right)+3\left[\frac{1}{d-3}-\Log{\left(\mu_s^2 r^2\right)}\right]\right\}\\
 & \nm
+\frac{G M^2 \nu \, r}{4} \left[4 (4+\nu )+\left(-3+24 \nu -10 \nu ^2\right)\left(\mathbf{n} \cdot \mathbf{v} \right)^2 +\left(17-14 \nu -26 \nu ^2\right)\mathbf{v}^2\right]\\
&
+\frac{M \nu \, r^2 \mathbf{v}^2}{8}\left[4-12 \nu +\left(3-23 \nu +47 \nu ^2\right)\mathbf{v}^2\right]\biggr\}\,,
\end{align}
respectively. It is straightforward to check that, upon taking the second-order time derivative and substituting the ($d$-dimensional) equations of motion, that the two expressions agree with each other.

\bibliography{Ref_NNNLO.bib}

\end{document}